\documentclass[3p,onecolumn,sort&compress]{elsarticle}
\usepackage[utf8]{inputenc}
\usepackage[english]{babel}
\usepackage{amssymb}
\usepackage{amsmath}
\usepackage{amsfonts}
\usepackage{url}
\usepackage{hyperref}
\usepackage{physics}
\usepackage{bm}
\usepackage{graphicx}
\usepackage{color}
\usepackage{subcaption}
\usepackage{tikz}
\usepackage{pgfplots}
\usepackage{multirow}
\usepackage{mdframed}

\pgfplotsset{compat=1.16}

\usepackage[normalem]{ulem}

\makeatletter
\def\ps@pprintTitle{%
 \let\@oddhead\@empty
 \let\@evenhead\@empty
 \def\@oddfoot{}%
 \let\@evenfoot\@oddfoot}
\makeatother

\graphicspath{ {./img/} }

\newcommand{\E}[1]{\cdot 10^{#1}}

\newcommand{\mrs}{\mathrm{s}}
\newcommand{\mre}{\mathrm{e}}
\newcommand{\mrn}{\mathrm{n}}
\newcommand{\mrp}{\mathrm{p}}
\newcommand{\typ}{\mathrm{typ}}
\newcommand{\tot}{\mathrm{tot}}
\newcommand{\SR}{\mathrm{SR}}
\newcommand{\SEI}{\mathrm{SEI}}
\newcommand{\Li}{\mathrm{Li}}

\DeclareMathOperator{\arcsinh}{arcsinh}

\usepackage{todonotes}

\newcommand{\rv}[1]{#1}
\newcommand{\rvv}[1]{#1}

\title{\rv{A Single Particle Model with Electrolyte and Side Reactions for degradation of lithium-ion batteries}}
\author[1,2,3]{Ferran Brosa Planella\corref{cor}}
\author[2,3]{W. Dhammika Widanage}

\address[1]{Mathematics Institute, University of Warwick, Gibbet Hill Road, Coventry, CV4 7AL, United Kingdom}
\address[2]{WMG, University of Warwick, Gibbet Hill Road, Coventry, CV4 7AL, United Kingdom}
\address[3]{The Faraday Institution, Harwell Campus, Didcot, OX11 0RA, United Kingdom}

\cortext[cor]{Corresponding author: Ferran.Brosa-Planella@warwick.ac.uk}

\begin{document}


\begin{abstract}
Battery degradation, which is the reduction of performance over time, is one of the main roadblocks to the wide deployment of lithium-ion batteries. Physics-based models, such as those based on the Doyle-Fuller-Newman model, are invaluable tools to understand and predict such phenomena. However, these models are often too complex for practical applications, so reduced models are needed. In this article we introduce the Single Particle Model with electrolyte and Side Reactions, a reduced model with electrochemical degradation which has been formally derived from the Doyle-Fuller-Newman model with Side Reactions using asymptotic methods. The reduced model has been validated against the full model for three scenarios (solid-electrolyte interphase growth, lithium plating, and both effects combined) showing similar accuracy at a much lower computational cost. The implications of the results are twofold: the reduced model is simple and accurate enough to be used in most real practical applications, and the reduction framework used is robust so it can be extended to account for further degradation effects.
\end{abstract}

\begin{keyword}
lithium-ion batteries \sep model reduction \sep Single Particle Model \sep lithium plating \sep solid-electrolyte interphase growth \sep asymptotic methods
\end{keyword}

\maketitle



\section{Introduction}



\rv{Addressing} climate emergency is one of the biggest challenges we face as society, and lithium-ion batteries are called to play a central role in the transition to a more sustainable future, in particular in the areas of transport electrification and off-grid energy storage. One of the main roadblocks for the wide deployment of batteries, especially in electric vehicles, is the decrease in the energy and power output as batteries age, known as degradation. Understanding and reducing such degradation would be extremely useful to extend the operation time of batteries and enable more efficient repurposing and recycling at the end of their life.

As described in \cite{Birkl2017,Edge2021}, degradation is caused by different mechanisms (or combinations thereof), which are triggered by multiple phenomena (e.g. high charge/discharge rates, extreme temperatures). These mechanisms contribute to the degradation modes: loss of active material (in each electrode), loss of lithium inventory, stoichiometric drift and impedance change \cite{Edge2021}. Ultimately, these modes manifest as a fade in the battery capacity and power outputs. According to \cite{Edge2021}, the main degradation mechanisms are solid-electrolyte interphase (SEI) growth and lithium plating (in the negative electrode), structural change and decomposition (in the positive electrode), and particle fracture (in both electrodes). Moreover, the mechanisms are not independent from each other, and positive feedback loops between various mechanisms have been reported \cite{Edge2021,Reniers2019}. 

Mathematical models are an extremely valuable tool to understand, predict and eventually reduce degradation via smart control of batteries. However, these models are very expensive from the computational point of view, which makes them unsuitable for many practical applications, especially those requiring a large number of simulations such as battery design and control. Therefore, there is a clear need for fast (yet accurate) models that capture battery degradation.


There are various approaches to battery modelling but here we focus our attention on physics-based continuum models \cite{Howey2020}. Physics-based continuum models treat the battery as a continuum material (i.e. they do not consider the atomistic level) and describe their behaviour by using conservation laws and constitutive relations to describe the transport of conserved quantities. These typically yield a coupled system of differential equations that needs to be solved numerically in order to determine any variable of interest. For a detailed review on continuum models for batteries, we refer the reader to the review article by Brosa Planella et al. \cite{BrosaPlanella2022}. Continuum models can include a wide range of physics but the cornerstone are electrochemical models, which describe the transport of lithium in the battery. There are different models in this category, with different levels of complexity, as reviewed in \cite{BrosaPlanella2022}. The most popular ones are the Doyle-Fuller-Newman model (DFN) introduced in \cite{Doyle1993,Fuller1994} and Single Particle Models (SPM), which are simpler. The latter can be defined either with electrolyte dynamics (e.g. \cite{BrosaPlanella2021,Marquis2019}) or without (e.g. \cite{Bizeray2019}). Any further effects, such as thermal and degradation, need to be coupled to an electrochemical model. In this article we focus our attention on two degradation mechanisms: SEI growth and lithium plating. These are probably the two most common degradation mechanisms, and they have been widely reported in the literature. For more information, we direct the reader to the review article by Edge et al. on battery degradation \cite{Edge2021}, and its companion article by O'Kane et al. on how to model it \cite{OKane2022}. 

The SEI is a passivation layer deposited at the surface of the electrode particles and which behaves like a solid electrolyte. This layer is created by the decomposition of the liquid electrolyte when it operates below its stability voltage window. During the formation phase of a newly assembled battery, the SEI layer is grown in a controlled manner. This causes a decrease in the battery capacity, but significantly reduces further reactions. However, as the battery ages, this layer grows again causing power and capacity fade. In the literature, we find three main approaches to physics-based modelling of SEI growth: density functional theory (DFT) models, continuum models and zero-dimensional models \cite{Marquis2020}. DFT models describe the processes at atomistic level and are extremely complex from the computational point of view \cite{Benitez2017,Shi2012}. Continuum models describe the SEI layer at a continuum level and account for the electron, ion and interstitial transport across this layer \cite{Marquis2020,Christensen2004,Colclasure2011}. Zero-dimensional models can be seen as a reduced version of continuum models in which the whole behaviour of the film is encapsulated by a boundary condition at the surface of the electrode particles. These are simpler models that can be easily coupled to the electrochemical models. As discussed in \cite{Marquis2020}, there are multiple instances of such models depending on the included physics (e.g. electron tunnelling \cite{Li2015}, interstitials transport \cite{Single2018}, and solvent diffusion with reactions \cite{Liu2014,Pinson2013,Safari2009}) or the limiting processes (e.g. solvent diffusion limited \cite{Ploehn2004}, electron migration limited \cite{Peled1979}, and reaction limited \cite{Ramadass2004}).

Lithium plating is a reaction in which lithium ions deposit on the electrode particles surface instead of intercalating into them. This reaction is reversible, but plated lithium is prone to react and form SEI and, in turn, the SEI growth can electrically isolate the plated lithium, leading to a loss of lithium inventory. Lithium plating can also lead to dendrite growth which can cause an internal short circuit in the battery and is exacerbated by fast charging, so in recent years this degradation mechanism has received a lot of attention \cite{Liu2016,Waldmann2018}. From the modelling point of view, lithium plating has been mostly described by zero-dimensional models. The first DFN model with lithium plating was introduced by Arora et al. \cite{Arora1999}. This initial model assumed a constant exchange current density, and a dependency on the lithium ion and plated lithium concentration was introduced first in \cite{Wood2016} for lithium metal electrodes, and later in \cite{Yang2018} for porous electrodes. It is worth noting that, in the literature, we find models assuming asymmetric \cite{Arora1999,Yang2018} and symmetric plating reactions \cite{Wood2016,OKane2020}, the latter being simpler.

Despite being fundamentally different from the chemical point of view, these two mechanisms are very similar from \rv{a} modelling point of view (they are both side reactions) and that is why they are often considered together, including in the present article. In recent years, there has been a surge in the interest of modelling SEI growth and lithium plating together. For example, the model in \cite{Yang2017} combines \rv{the SEI model \cite{Safari2009} (solvent diffusion with reactions)} with an irreversible plating model similar to \cite{Arora1999}, and its analysis has shown that the interactions between SEI growth and lithium plating trigger nonlinear aging, also known as the \emph{knee} (see \cite{Attia2022} for a detailed review on the knee). Similar results have been reported in \cite{Atalay2020} (which includes two types of SEI reactions) and \cite{Keil2020} (which includes partially reversible lithium plating). The latter showed that at least a fraction of irreversible lithium plating is required in order to trigger the knee. In all three cases, the models are validated with experimental data, showing good agreement. Finally, in \cite{OKane2022}, SEI growth and lithium plating are also coupled with particle cracking and loss of active material, showcasing a very diverse range of behaviours depending on the cycling profile. The authors highlight the need for further studies and more accurate parameter sets in order to discover potential new degradation regimes.


As explained earlier, degradation models are coupled to electrochemical models, to which they add complexity. If the standard DFN model is already complex enough for many applications, adding degradation effects only increases this complexity \rv{\cite{Sikha2004,Ning2006}}. For this reason, researchers have explored coupling degradation effects to simpler models, such as Single Particle Models \rv{\cite{Safari2009,Lin2018,Pang2019,Yu2022}}. However, this is typically done in an \emph{ad hoc} manner, which often leads to inconsistencies. For example, in \rv{\cite{Lin2018,Pang2019,Yu2022}} the loss of interfacial current to the SEI reaction is not taken into account, and thus the total amount of lithium is not conserved. Therefore, it is clear that such simplified models for battery degradation should be rigorously derived to ensure they are consistent with the physical principles of the DFN model. Asymptotic methods provide a robust framework to simplify complex models in a systematic manner. Such methods have already been applied to battery models such as the DFN model to derive Single Particle Models and other reduced models (e.g. \cite{BrosaPlanella2021,Marquis2019,Moyles2019}) but, to the best of our knowledge, they have not yet been applied to derive models accounting for degradation. 

\rvv{We will use standard asymptotic methods \cite{Bender1999,Hinch1991}, also known as perturbation methods, to simplify the model based on \emph{a priori} assumptions about the size of certain dimensionless parameters. Asymptotic methods offer the advantages of being generic, systematic and flexible. Therefore, we can use them to simplify battery degradation models while ensuring that the reduced models are physically consistent with the full models.}

In this article we present the Single Particle Model with electrolyte and Side Reactions (SPMe+SR), a simplified model for electrochemical battery degradation (Section \ref{sec:SPMe+SR}). The SPMe+SR is formally derived from the Doyle-Fuller-Newman model with Side Reactions (DFN+SR) using asymptotic methods (Section \ref{sec:derivation_SPMe_SR}), \rv{and based on the assumptions of small overpotentials and weak side reactions}.  The SPMe+SR is generic enough to account for a wide range of side reactions \rv{and operating conditions}, and is found to offer similar accuracy to the DFN+SR while being much simpler (see Section \ref{sec:results}). This means that the SPMe+SR is easier to analise and faster to simulate, making it suitable to a wider range of applications than the DFN+SR \rv{such as real-time control, prediction of the remaining useful life, and design optimisation.}

\rv{To the best of our knowledge, the results presented in this article are the first instance of a Single Particle Model with side reactions which is physically consistent and has been systematically derived from the full DFN+SR model.} The derivation method presented here can be applied to models accounting for other degradation mechanisms, and thus it is a stepping stone towards a robust and unified framework for reduced models for battery degradation.

\section{Single Particle Model with electrolyte and Side Reactions (SPMe+SR)}\label{sec:SPMe+SR}

In this section we present the reduced model for battery degradation that we named Single Particle Model with electrolyte and Side Reactions (SPMe+SR). This model belongs to the family of Single Particle Models with electrolyte dynamics (see \cite{BrosaPlanella2022,BrosaPlanella2021} for detailed discussions on this model family), and models a single averaged (or representative) particle to describe the behaviour of all the particles within each electrode. The SPMe+SR presented here is an electrochemical model accounting for degradation in the negative electrode caused by a side reaction (i.e. an undesired reaction that consumes lithium ions and produces new material that blocks the pores in the electrode), but it could be very easily extended to account for side reactions in the positive electrode as well. The key advantage of the SPMe+SR model, with respect to the DFN+SR, is that it only requires solving for the \rv{electrolyte and averaged particle concentrations}, and the electrode porosity, while all the other variables of interest (such as the potentials and currents) can be calculated from explicit expressions. \rv{In comparison, the DFN+SR requires solving for the electrolyte and particle concentrations (the latter as a function of radius and electrode thickness), electrode and electrolyte potentials, and electrode porosity. For a detailed comparison between SPMe and DFN models we refer the reader to \cite{BrosaPlanella2022}.}

\begin{figure}
    \centering
    \includegraphics[scale=1]{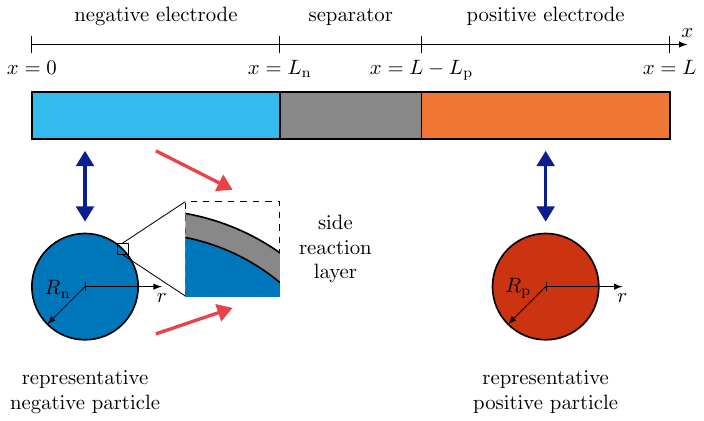}
    \caption{Geometry of the SPMe+SR model. The upper layer represents the porous electrodes ($0 \leq x \leq L$) in which we solve the electrolyte equation. For each electrode we have a representative particle ($0 \leq r \leq R_k$, for $k \in \{\mrn, \mrp \}$). In addition, in the negative electrode we also have the side reaction layer. The blue arrows represent the intercalation reaction (reversible), while the red arrows represent the side reaction (typically irreversible).}
    \label{fig:geometry}
\end{figure}

The geometry of the model is as shown in Figure~\ref{fig:geometry}. The two averaged electrode particles are defined in the domain $0 \leq r \leq R_k$ (for $k \in \{\mrn,\mrp\}$), where the subscripts $\mrn$ and $\mrp$ denote the negative and positive electrode, respectively. The domain for the electrolyte is $0 \leq x \leq L$, and includes the negative electrode ($0 \leq x \leq L_\mrn$), the separator ($L_\mrn \leq x \leq L - L_\mrp$), and the positive electrode ($L - L_\mrp \leq x \leq L$). In each electrode particle we solve a partial differential equation (PDE) for the lithium concentration \eqref{eq:SPMe_cn}-\eqref{eq:SPMe_cp}, while in the electrolyte domain we solve a PDE for the ion concentration in the electrolyte \eqref{eq:SPMe_ce} and one for the porosity \eqref{eq:SPMe_eps_n} (note that the latter only applies to the electrolyte in the negative electrode subdomain). The key advantage of the SPMe+SR is that all the PDEs involved are time dependent and thus, after spatial discretisation (and assuming that the current is the input), they become a system of ordinary differential equations (ODEs) rather than a system of differential-algebraic equations (DAEs) which would be much harder to solve numerically. The main difference with the standard SPMe is that for the SPMe+SR the various PDEs are coupled together through the side reaction. This is similar to what occurs with the thermal SPMe derived in \cite{BrosaPlanella2021}, so we refer the keen reader to that article for more details on SPMe.

The SPMe+SR is derived from the DFN+SR using asymptotic methods (see Section \ref{sec:derivation_SPMe_SR} for details), and this analysis is based on certain assumptions that define the range of validity of the SPMe+SR. The two assumptions are: small deviations from the open-circuit potential \rv{(i.e. overpotentials)} and weak side reaction. The first one is reasonable for low to moderate C-rates, and would only break down at high C-rates. However, in practice, other limiting phenomena occur before such regimes are reached, such as electrolyte depletion. \rv{When depletion occurs, the behaviour of the electrode particles is very different depending on whether they are in the depleted region or not, and the single particle model is no longer valid.} The second assumption is reasonable because, if the side reaction was not small compared to the intercalation reaction, the battery would show a Coulombic efficiency much lower than what is observed in practice \cite{Yang2017}. Therefore, the two assumptions are reasonable for a broad range of operating conditions. Having discussed its range of validity, we can now define the SPMe+SR.

\paragraph{Particle equations} The equation for the lithium concentration in the negative averaged (or representative) particle reads
\begin{subequations}\label{eq:SPMe_cn}
\begin{align}
\pdv{\bar c_\mrn}{t} &= \frac{1}{r^2} \pdv{}{r} \left(r^2 D_\mrn(\bar c_\mrn) \pdv{\bar c_\mrn}{r} \right), & \quad \text{ in } 0 < r < R_\mrn, \label{eq:SPMe_cn_a}\\
\pdv{\bar c_\mrn}{r} &= 0, & \quad \text{ at } r = 0,\\
- D_\mrn (\bar c_\mrn) \pdv{\bar c_\mrn}{r} &= \frac{1}{a_\mrn F} \left( \frac{i_\mathrm{app}}{L_\mrn} - \bar J_\SR \right), & \quad \text{ at } r = R_\mrn, \label{eq:SPMe_cn_c}\\
\bar c_\mrn &= c_{\mrn,\mathrm{init}}, & \quad \text{ at } t = 0,
\end{align}
\end{subequations}
where
\begin{subequations}
\begin{align}
    \bar J_\SR &= \frac{1}{L_\mrn} \int_0^{L_\mrn} J_\SR \dd x, \label{eq:SPMe_J_SR_av}\\
    J_\SR &= - a_\mrn j_{\SR} \exp \left( - \alpha_\SR \frac{F}{R T}\left( \phi_{\mrn} - \phi_\mre - U_\SR - \frac{i_\mathrm{app} L_{\mathrm{f},\mrn}}{L_\mrn a_\mrn \Sigma_{\mathrm{f},\mrn}} \right) \right).
\end{align}
\end{subequations}

Here the variable to solve for is the concentration in the negative averaged particle $\bar c_\mrn (r, t)$, and the parameters are the diffusivity $D_\mrn(\bar c_\mrn)$, the surface area per unit volume \rv{of the active material particles (without the side reaction layer)} $a_\mrn$, the Faraday constant $F$, the applied current density $i_\mathrm{app}(t)$, the thickness of the negative electrode $L_\mrn$, and the initial concentration of the particle $c_{\mrn,\mathrm{init}}$. The side reaction current density (per unit volume) is $J_\SR$, and the parameters that define it are the exchange current density $j_\SR$ (which may be a function of the variables of the problem, see Section \ref{sec:SR_models}), the transfer coefficient $\alpha_\SR$, the gas constant $R$, the temperature $T$, the open-circuit potential $U_\SR$, and the film conductivity $\Sigma_{\mathrm{f},\mrn}$. The electrode and electrolyte potentials ($\phi_\mrn$ and $\phi_\mre$, respectively) and the film thickness $L_{\mathrm{f},\mrn}$ are quantities that depend on the variables of the model and are computed from the expressions \eqref{eq:SPMe_film_thickness}-\eqref{eq:SPMe_potentials}. Note that we use the overhead bar to define the average over the thickness of the corresponding electrode, as defined in \eqref{eq:SPMe_J_SR_av}. The definition of the concentration $\bar c_\mrn$ as an averaged quantity arises from the definition of the Single Particle Model (see Section \ref{sec:derivation_SPMe_SR}).

The equation for the lithium concentration in the positive averaged (or representative) particle equation reads
\begin{subequations}\label{eq:SPMe_cp}
\begin{align}
\pdv{\bar c_\mrp}{t} &= \frac{1}{r^2} \pdv{}{r} \left(r^2 D_\mrp(\bar c_\mrp) \pdv{\bar c_\mrp}{r} \right), & \quad \text{ in } 0 < r < R_\mrp,\\
\pdv{\bar c_\mrp}{r} &= 0, & \quad \text{ at } r = 0,\\
- D_\mrp (\bar c_\mrp) \pdv{\bar c_\mrp}{r} &= - \frac{1}{a_\mrp F} \frac{i_\mathrm{app}}{L_\mrp}, & \quad \text{ at } r = R_\mrp,\\
\bar c_\mrp &= c_{\mrp,\mathrm{init}}, & \quad \text{ at } t = 0.
\end{align}
\end{subequations}
Here the variable to solve for is the concentration in the positive averaged particle $\bar c_\mrp (r, t)$, and the parameters are analogous to those in the negative electrode particle.

\paragraph{Electrolyte equation} The equation for the ion concentration in the electrolyte reads
\begin{subequations}\label{eq:SPMe_ce}
\begin{align}
    \pdv{}{t} \left( \varepsilon c_{\mre} \right) &=  \pdv{}{x} \left( D_\mre (c_{\mre}) \mathcal{B}(x, t) \pdv{c_{\mre}}{x} + \frac{(1 - t^+(c_{\mre}))}{F} i_{\mre} \right), & \text{ in } 0 \leq x \leq L,\\
    \pdv{c_{\mre}}{x} &= 0, & \text{ at } x = 0,L,\\
    c_{\mre} &= c_{\mre,\mathrm{init}}, & \text{ at } t = 0.
\end{align}
\end{subequations}

Here the variable to solve for is the lithium ion concentration in the electrolyte $c_\mre (x, t)$, and the parameters are the diffusivity $D_\mre (c_\mre)$, the transport efficiency (also known as inverse MacMullin number) $\mathcal{B}(x, t)$, the transference number $t^+(c_\mre)$, and the initial concentration $c_{\mre,\mathrm{init}}$. The current in the electrolyte $i_\mre$ is defined as in \eqref{eq:SPMe_ie}, and the porosity $\varepsilon(x, t)$ is calculated from \eqref{eq:SPMe_porosity}.

\paragraph{Porosity equation} The equation for the the porosity in the negative electrode is
\begin{equation}\label{eq:SPMe_eps_n}
    \pdv{\varepsilon_\mrn}{t} = \frac{M_\SR}{n_\SR \rho_\SR F} J_\SR,
\end{equation}
where $M_\SR$ and $\rho_\SR$ are the molar weight and density of the side reaction material, respectively, and $n_\SR$ the number of electrons involved in the side reaction. The porosity across the battery is defined piecewise as
\begin{equation}\label{eq:SPMe_porosity}
    \varepsilon (x, t) = 
    \begin{cases}
    \varepsilon_\mrn (x, t), & \text{ if } 0 \leq x < L_\mrn,\\
    \varepsilon_\mrs (x), & \text{ if } L_\mrn \leq x < L - L_\mrp,\\
    \varepsilon_\mrp (x), & \text{ if } L - L_\mrp \leq x \leq L.
    \end{cases}
\end{equation}
In this case, we define the side reaction only in the negative electrode and thus the porosity in the separator and positive electrode remain constant in time, but the model could be easily extended to account for a side reaction and porosity variation in the positive electrode as well.

The thickness of the side reaction film can be calculated as
\begin{equation}\label{eq:SPMe_film_thickness}
    L_{\mathrm{f},k} = L_{\mathrm{f},k, \mathrm{init}} - \frac{1}{a_k} (\varepsilon_k - \varepsilon_{k, \mathrm{init}}),
\end{equation}
and note that this reduces to a trivial $L_{\mathrm{f},k} = L_{\mathrm{f},k, \mathrm{init}}$ in the case where the porosity does not change (i.e. the film does not grow).

After solving equations \eqref{eq:SPMe_cn}, \eqref{eq:SPMe_cp}, \eqref{eq:SPMe_ce} and \eqref{eq:SPMe_eps_n} to determine $\bar c_\mrn$, $\bar c_\mrp$, $\bar c_\mre$ and $\varepsilon$, we can compute any other variable of interest from explicit expressions.

\paragraph{Currents in the electrodes and electrolyte} The currents in the electrodes and electrolyte are defined as follows. In the negative electrode ($0 \leq x \leq L_\mrn$) the current is
\begin{subequations}
\begin{equation}
    i_\mrn(x, t) = \frac{i_\mathrm{app}}{L_\mrn} \left(L_\mrn - x \right),
\end{equation}
while in the positive electrode ($L - L_\mrp \leq x \leq L$) it is
\begin{equation}
    i_\mrp(x, t) = \frac{i_\mathrm{app}}{L_p} \left(x - (L - L_\mrp) \right).
\end{equation}
The current in the electrolyte is
\begin{align}\label{eq:SPMe_ie}
    i_{\mre}(x, t) &= \begin{cases}
    \frac{i_\mathrm{app}}{L_\mrn}x, & \text{ if } 0 \leq x < L_\mrn,\\
    i_\mathrm{app}, & \text{ if } L_\mrn \leq x < L - L_\mrp,\\
    \frac{i_\mathrm{app}}{L_\mrp} (L - x), & \text{ if } L - L_\mrp \leq x \leq L.
    \end{cases}
\end{align}
\end{subequations}

\paragraph{Potentials in the electrodes and electrolyte} The potentials in the electrodes and electrolyte are defined as follows. In the negative and positive electrodes, respectively, the potentials are
\begin{subequations}\label{eq:SPMe_potentials}
\begin{align}
    \phi_\mrn(x, t) &= U_\mrn \left( \left. \bar c_\mrn \right|_{r = R_\mrn} \right) - \frac{i_\mathrm{app} (2 L_\mrn - x) x}{2 L_\mrn \sigma_\mrn} + \frac{i_\mathrm{app} L_\mrn}{3 \sigma_\mrn} + \frac{i_\mathrm{app} \bar L_{\mathrm{f},\mrn}}{L_\mrn a_\mrn \sigma_{\mathrm{f},\mrn}} \nonumber \\
    & \quad - \frac{1}{L_\mrn} \int_0^{L_\mrn} \int_0^x \frac{i_{\mre}(s,t) \dd s}{\sigma_\mre (c_{\mre}(s,t)) \mathcal{B}(s, t)} \dd x \nonumber \\
    & \quad + \frac{2 R T}{F} \frac{1}{L_\mrn} \int_0^{L_\mrn} \int_0^x (1 - t^+(c_{\mre}(s, t))) \left( 1 + \pdv{f_\pm}{c_\mre} \right) \pdv{\log c_{\mre}(s,t)}{s} \dd s \dd x \nonumber \\
    & \quad + \frac{2 R T}{F} \frac{1}{L_\mrn} \int_0^{L_\mrn} \arcsinh\left(\frac{i_\mathrm{app}}{a_\mrn L_\mrn j_{\mrn}} \right) \dd x, \label{eq:phi_n_dim} \\
    \phi_\mrp (x, t) &= U_\mrp \left( \left. \bar c_\mrp \right|_{r = R_\mrp} \right) + \frac{i_\mathrm{app} (2 (L - L_\mrp) - x) x}{2 L_\mrp \sigma_\mrp} - \frac{i_\mathrm{app}(2 L_\mrp^2 - 6 L_\mrp + 3)}{6 L_\mrp \sigma_\mrp} - \frac{i_\mathrm{app} \bar L_{\mathrm{f},\mrp}}{L_\mrp a_\mrp \sigma_{\mathrm{f},\mrp}} \nonumber \\
    & \quad - \frac{1}{L_\mrp} \int_{L - L_\mrp}^L \int_0^x \frac{i_{\mre}(s,t) \dd s}{\sigma_\mre (c_{\mre}(s,t)) \mathcal{B}(s, t)} \dd x \nonumber \\
    & \quad + \frac{2 R T}{F} \frac{1}{L_\mrp} \int_{L - L_\mrp}^L \int_0^x (1 - t^+(c_{\mre}(s, t))) \left( 1 + \pdv{f_\pm}{c_\mre} \right) \pdv{\log c_{\mre}(s,t)}{s} \dd s \dd x \nonumber \\ 
    & \quad - \frac{2 R T}{F} \frac{1}{L_\mrp} \int_{L - L_\mrp}^L \arcsinh\left(\frac{i_\mathrm{app}}{a_\mrp L_\mrp j_{\mrp}} \right) \dd x, \label{eq:phi_p_dim} 
\end{align}
while in the electrolyte the potential is
\begin{equation}\label{eq:phi_e_dim} 
    \phi_\mre (x, t) = - \int_0^x \frac{i_{\mre}}{\sigma_\mre (c_{\mre}(s, t)) \mathcal{B}(s, t)} \dd s + \frac{2 R T}{F} \int_0^x (1 - t^+(c_{\mre}(s, t))) \left( 1 + \pdv{f_\pm}{c_\mre} \right) \pdv{\log c_{\mre}(s,t)}{s} \dd s,
\end{equation}
\end{subequations}
with the intercalation exchange current density defined as
\begin{align}
    j_k &= m_k \left. \sqrt{c_{\mre} \bar c_{k} \left(c_k^{\max} - \bar c_{k} \right)} \right|_{r = R_k}.
\end{align}

Here the new parameters introduced (for $k \in \{\mrn,\mrp\}$) are the open-circuit potentials of the electrodes $U_k (c_k)$, the electrode electronic conductivities $\sigma_k$, the electrolyte conductivity $\sigma_\mre (c_\mre)$, the thermodynamic factor $\left( 1 + \pdv{f_\pm}{c_\mre} \right)$, the intercalation reaction constant $m_k$, and the maximum concentration in the particles $c_k^{\max}$. \rv{Even though the thermodynamic factor is written as $\left( 1 + \pdv{f_\pm}{c_\mre} \right)$, it is usually used as a single parameter directly fitted to experimental data \cite{Landesfeind2019}.} \rv{Note that we have assumed the Butler-Volmer reaction to have transfer coefficients equal to 0.5, but the results can be easily generalised to non-symmetric reactions \cite{BrosaPlanella2022}.}

\paragraph{Terminal voltage} The terminal voltage can be calculated from the electrode potentials as
\begin{equation}
    V(t) = \left. \phi_\mrp \right|_{x = L} - \left. \phi_\mrn \right|_{x = 0}.
\end{equation}
Using the definitions for the potentials \eqref{eq:phi_n_dim}-\eqref{eq:phi_p_dim}, we can be write the terminal voltage as
\begin{equation}
    V(t) = U_\mathrm{eq} + \eta_\mathrm{r} + \eta_\mre + \Delta \phi_\mre + \Delta \phi_\mrs + \Delta \phi_\mathrm{f},
\end{equation}
where
\begin{equation}
\begin{aligned}
U_\mathrm{eq} &= U_\mrp \left( \left. \bar c_\mrp \right|_{r = R_\mrp} \right) - U_\mrn \left( \left. \bar c_\mrn \right|_{r = R_\mrn} \right),\\
\eta_\mathrm{r} &= - \frac{2 R T}{F} \left(\frac{1}{L_\mrp} \int_{L - L_\mrp}^L \arcsinh\left(\frac{i_\mathrm{app}}{a_\mrp L_\mrp j_{\mrp}} \right) \dd x + \frac{1}{L_\mrn} \int_0^{L_\mrn} \arcsinh\left(\frac{i_\mathrm{app}}{a_\mrn L_\mrn j_{\mrn}} \right) \dd x \right),\\
\eta_\mre &= \frac{2 R T}{F} \left( \frac{1}{L_\mrp} \int_{L - L_\mrp}^L \int_0^x (1 - t^+(c_{\mre}(s, t))) \pdv{\log c_{\mre}(s,t)}{s} \dd s \dd x \right.\\
& \quad \left.- \frac{1}{L_\mrn} \int_0^{L_\mrn} \int_0^x (1 - t^+(c_{\mre}(s, t))) \pdv{\log c_{\mre}(s,t)}{s} \dd s \dd x \right),\\
\Delta \phi_\mre &= - \frac{1}{L_\mrp} \int_{L - L_\mrp}^L \int_0^x \frac{i_{\mre}(s,t) \dd s}{\sigma_\mre (c_{\mre}(s,t)) \mathcal{B}(s, t)} \dd x + \frac{1}{L_\mrn} \int_0^{L_\mrn} \int_0^x \frac{i_{\mre}(s,t) \dd s}{\sigma_\mre (c_{\mre}(s,t)) \mathcal{B}(s, t)} \dd x,\\
\Delta \phi_\mrs &= - \frac{i_\mathrm{app}}{3} \left(\frac{L_\mrp}{\sigma_\mrp} + \frac{L_\mrn}{\sigma_\mrn} \right),\\
\Delta \phi_\mathrm{f} &= - i_\mathrm{app} \left( \frac{\bar L_{\mathrm{f},\mrp}}{L_\mrp a_\mrp \sigma_{\mathrm{f},\mrp}} + \frac{\bar L_{\mathrm{f},\mrn}}{L_\mrn a_\mrn \sigma_{\mathrm{f},\mrn}} \right).
\end{aligned}
\end{equation}

Each term in the voltage expression has a physical interpretation: $U_\mathrm{eq}$ is the equilibrium (or open-circuit) potential of the cell, $\eta_\mathrm{r}$ is the intercalation reaction overpotential, $\eta_\mre$ is the electrolyte concentration overpotential, $\Delta \phi_\mre$ are the Ohmic losses in the electrolyte, $\Delta \phi_\mrs$ are the Ohmic losses in the electrode, and $\Delta \phi_\mathrm{f}$ are the Ohmic losses in the interface film.

\subsection{Examples of SR models}\label{sec:SR_models}
The SPMe+SR is generic enough to account for a wide range of side reactions (or combinations thereof). In Section \ref{sec:results} we have simulated three cases: SEI growth, lithium plating, and their combined effect. Both SEI and plating models are taken to match those in \cite{Yang2017}, so they are \rv{an SEI growth model with solvent diffusion and reactions,} and an irreversible model for lithium plating. \rv{Note that it is out of the scope of this article to discuss the validity of such assumptions, and this choice of side reaction models is merely to illustrate the capabilities of the SPMe+SR. The model would also work with other choices of SEI and plating models.}

For SEI, we define $j_\SR = j_\SEI$ where
\rv{\begin{equation}\label{eq:j_SEI}
\begin{aligned}
    j_\SEI &= F k_\SEI c_\SEI,\\
    c_\SEI &= c_{\SEI,\mathrm{init}} + \frac{j_\SEI L_\mathrm{f}}{F D_\SEI}.
\end{aligned}
\end{equation}
Here $c_\SEI$ is the concentration of the SEI material (e.g. ethylene carbonate, also called EC)}, $k_\SEI$ is the reaction rate of the SEI reaction, $c_{\SEI,\mathrm{init}}$ is the initial SEI concentration, and $D_\SEI$ is the solvent diffusivity. More details on this model can be found in \cite{Safari2009,Yang2017}.

For lithium plating we define $j_\SR = j_\mathrm{Li}$ where
\begin{equation}\label{eq:j_Li}
    j_\Li = F k_\Li c_\mre,
\end{equation}
and $k_\Li$ is the lithium plating reaction rate. This model is slightly different to that in \cite{Yang2017} as the plating exchange current density depends on the ion concentration in the electrolyte. More details of this model can be found in \cite{OKane2020}.

Finally, both models can be combined to account for simultaneous SEI growth and lithium plating. In this case, we need to define $J_\mathrm{SR} = J_\SEI + J_\mathrm{Li}$, where 
\begin{equation}
\begin{aligned}
    J_\SEI &= - a_\mrn j_{\SEI} \exp \left( - \alpha_\SEI \frac{F}{R T}\left( \phi_{\mrn} - \phi_\mre - U_\SEI - \frac{i_\mathrm{app} L_{\mathrm{f},\mrn}}{L_\mrn a_\mrn \Sigma_{\mathrm{f},\mrn}} \right) \right),\\
    J_\Li &= - a_\mrn j_{\Li} \exp \left( - \alpha_\Li \frac{F}{R T}\left( \phi_{\mrn} - \phi_\mre - U_\Li - \frac{i_\mathrm{app} L_{\mathrm{f},\mrn}}{L_\mrn a_\mrn \Sigma_{\mathrm{f},\mrn}} \right) \right),
\end{aligned}
\end{equation}
and $j_\SEI$ and $j_\Li$ are defined from \eqref{eq:j_SEI}-\eqref{eq:j_Li}. We now need to be careful on how to define the porosity variation as both reactions will contribute to it. In this case it is easier to define the thickness of the films created by each reaction as
\begin{equation}\label{eq:film_equations_SEI_plating}
\begin{aligned}
    \pdv{L_{\mathrm{f},\SEI}}{t} &= - \frac{M_\SEI}{n_\SEI \rho_\SEI F} J_\SEI,\\
    \pdv{L_{\mathrm{f},\Li}}{t} &= - \frac{M_\mathrm{Li}}{n_\mathrm{Li} \rho_\mathrm{Li} F} J_\mathrm{Li}.
\end{aligned}
\end{equation}
With the two thicknesses separately, we can define the total film thickness as $L_{\mathrm{f},\mrn} = L_{\mathrm{f},\SEI} + L_{\mathrm{f},\Li}$, and the film conductivity according to modelling assumptions of our choice. For the results in Section \ref{sec:results}, we assume that plated lithium is an ideal conductor and therefore the film Ohmic losses are caused by SEI only.

We can then retrieve the porosity from inverting \eqref{eq:SPMe_film_thickness}, which yields
\begin{equation}
        \varepsilon_k = \varepsilon_{k, \mathrm{init}} - a_k (L_{\mathrm{f},k} - L_{\mathrm{f},k, \mathrm{init}}).
\end{equation}
Note that taking the time derivative of this equation and substituting \eqref{eq:film_equations_SEI_plating} into it, we obtain the porosity equation stated in \cite{Yang2017}
\begin{equation}\label{eq:porosity_combined_SR}
    \pdv{\varepsilon_\mrn}{t} = \frac{M_\SEI}{n_\SEI \rho_\SEI F} J_\SEI + \frac{M_\mathrm{Li}}{n_\mathrm{Li} \rho_\mathrm{Li} F} J_\mathrm{Li}.
\end{equation}

\subsection{Conservation of lithium}
A fundamental feature of any battery model is to conserve lithium. In this section we prove that the SPMe+SR conserves the total amount of lithium. The total amount of lithium in the battery $N_\mathrm{tot}$ is given by
\begin{equation}
    N_\mathrm{tot} (t) = N_\mrn (t) + N_\mrp (t) + N_\mre (t) + N_\SR (t),
\end{equation}
where the total amounts of lithium in each domain are $N_\mrn$ for the negative electrode, $N_\mrp$ for the positive electrode, $N_\mre$ for the electrolyte and $N_\SR$ for the side reaction material. These quantities are defined as the integral of the corresponding concentration of lithium species over the domain they occupy, hence
\begin{equation}
\begin{aligned}
    N_\mrn (t) &= A L_\mrn \frac{3 \varepsilon_{\mrs, \mrn}}{R_\mrn^3} \int_0^{R_\mrn} r^2 \bar c_\mrn \dd r,\\
    N_\mrp (t) &= A L_\mrp \frac{3 \varepsilon_{\mrs, \mrp}}{R_\mrp^3} \int_0^{R_\mrp} r^2 \bar c_\mrp \dd r,\\
    N_\mre (t) &= A \int_0^L \varepsilon c_\mre \dd x,\\
    N_\SR (t) &= A \int_0^{L_\mrn} \frac{n_\SR \rho_\SR}{M_\SR} (\varepsilon_{\mrn,\mathrm{init}} - \varepsilon_\mrn) \dd x,
\end{aligned}
\end{equation}
where $\varepsilon_{\mrs,k}$ for $k \in \{\mrn,\mrp\}$ is the volume fraction of active material in each electrode, \rv{and $A$ is the current collector surface}.

pWe want to prove that $\dv{N_\mathrm{tot}}{t} = 0$, therefore we need to calculate the time derivatives for the total amount of lithium in each domain. For example, for the negative electrode we have
\begin{align}
    \dv{N_\mrn}{t} &= A L_\mrn \frac{3 \varepsilon_{\mrs, \mrn}}{R_\mrn^3} \int_0^{R_\mrn} r^2 \pdv{\bar c_\mrn}{t} \dd r,
\end{align}
and using the expressions in \eqref{eq:SPMe_cn_a} and \eqref{eq:SPMe_cn_c} we can rewrite it as
\begin{multline}
    \dv{N_\mrn}{t} = A L_\mrn \frac{3 \varepsilon_{\mrs, \mrn}}{R_\mrn^3} \int_0^{R_\mrn} \pdv{}{r} \left(r^2 D_\mrn (\bar c_\mrn) \pdv{\bar c_\mrn}{r} \right) \dd r = \left. A L_\mrn \frac{3 \varepsilon_{\mrs, \mrn}}{R_\mrn^3} r^2 D_\mrn (\bar c_\mrn) \pdv{\bar c_\mrn}{r} \right]_{r = 0}^{r = R_\mrn} \\
    = - A L_\mrn \frac{3 \varepsilon_{\mrs, \mrn}}{R_\mrn} \frac{1}{a_\mrn F} \left(\frac{i_\mathrm{app}}{L_\mrn} - \bar J_\SR \right)  = - \frac{A}{F} \left( i_\mathrm{app} - \bar J_\SR L_\mrn \right),
\end{multline}
where we have used that for spherical particles $a_k = \frac{3 \varepsilon_{\mrs,k}}{R_k}$, for $k \in \{\mrn,\mrp\}$.

Similarly, we can show that
\begin{equation}
\begin{aligned}
    \dv{N_\mrp}{t} &= \frac{A}{F} i_\mathrm{app},\\
    \dv{N_\mre}{t} &= 0,\\
    \dv{N_\SR}{t} &= - \frac{A}{F} \bar J_\SR L_\mrn,
\end{aligned}
\end{equation}
so the variation of the total amount of lithium is
\begin{equation}
    \dv{N_\mathrm{tot}}{t} = \frac{A}{F} \left( - i_\mathrm{app} + \bar J_\SR L_\mrn + i_\mathrm{app} - \bar J_\SR L_\mrn \right) = 0,
\end{equation}
hence the total amount of lithium is conserved.

\rv{Conservation of lithium is a very important feature of battery models which is often not ensured when the models are posed in an \emph{ad hoc} manner \cite{Lin2018,Pang2019,Yu2022}. For example, in \cite{Lin2018,Yu2022}, The boundary condition in the negative particles does not take into account the side reaction, and is simply}
\begin{equation}\label{eq:wrong_BC}
    - D_\mrn (\bar c_\mrn) \pdv{\bar c_\mrn}{r} = \frac{1}{a_\mrn F}\frac{i_\mathrm{app}}{L_\mrn}.
\end{equation}
\rv{Therefore, we find that the total amount of lithium is not conserved, as}
\begin{equation}
    \dv{N_\tot}{t} = \dv{N_\SR}{t} = - \frac{A}{F} \bar J_\SR L_\mrn > 0,
\end{equation}
\rv{so, in fact, the total amount of lithium in the system increases. In addition, the total amount of available lithium in the electrodes remains constant, so no capacity fade would be observed with this model.}

\rv{On the other hand, the model presented in \cite{Pang2019} also defines the negative particle boundary condition as \eqref{eq:wrong_BC}, but in addition it takes the source term of the electrolyte equation in the negative electrode to be}
\begin{equation}
    \frac{1}{F}\pdv{i_\mre}{x} = \frac{i_\mathrm{app}}{a_\mrn F L_\mrn} + J_\SR.
\end{equation}
\rv{Repeating the calculations on the total amount of lithium we find that}
\begin{equation}
    \dv{N_\tot}{t} = \dv{N_\mre}{t} + \dv{N_\SR}{t} = - \frac{A}{F} t^+ \bar J_\SR L_\mrn > 0,
\end{equation}
\rv{and again the total amount of lithium in the system increases. These inconsistencies highlight the importance of deriving the reduced models in a systematic manner to ensure that the conservation laws of the full model are preserved.}

\section{Derivation of the SPMe+SR}\label{sec:derivation_SPMe_SR}

In this section we derive the reduced SPMe+SR model from the full DFN+SR model presented in \ref{sec:DFN+SR}. We first state the rescaled dimensionless model (Section \ref{sec:rescaled_model}), which is the starting point of the asymptotic reduction in Section \ref{sec:asymptotic_reduction}. A summary of the dimensionless reduced model is presented in \ref{sec:model_summary}.

\subsection{Rescaled dimensionless model}\label{sec:rescaled_model}
We derive the SPMe+SR model from the isothermal DFN+SR, the dimensionless version of which is derived in \ref{sec:DFN+SR_ND}. The model has a set of dimensionless parameters which are defined in \eqref{eq:ND_parameters}. 
\rv{All the equations in this section are in dimensionless form.} The model for concentration in the particles reads
\begin{subequations}\label{eq:DFN+SR_ND_resc}
\begin{align}
\mathcal{C}_k \pdv{c_{k}}{t} &= \frac{1}{r^2} \pdv{}{r} \left(r^2 D_{k} \pdv{c_{k}}{r} \right), & \quad \text{ in } 0 < r < 1,\\
\pdv{c_{k}}{r} &= 0, & \quad \text{ at } r = 0,\\
- D_{k} \pdv{c_{k}}{r} &= \frac{\mathcal{C}_k}{\alpha_k \gamma_k} J_{k,\mathrm{int}}, & \quad \text{ at } r = 1,\\
c_{k} &= \mu_k, & \quad \text{ at } t = 0,
\end{align}
and the potential in each electrode is given by
\begin{align}
\pdv{i_{k}}{x} &= -J_k, \label{eq:DFN+SR_ND_resc_e}\\
i_{k} &= - \lambda \Sigma_{k} \pdv{\phi_{k}}{x}. \label{eq:DFN+SR_ND_resc_f}
\end{align}
The $x$ domain is defined to be $0 \leq x \leq \ell_\mrn$ if $k = \mrn$, and $1 - \ell_\mrp \leq x \leq 1$ if $k = \mrp$.

The electrolyte equations, which are defined in the domain $0 \leq x \leq 1$, are
\begin{align}
\mathcal{C}_\mre \gamma_\mre \pdv{}{t} \left( \varepsilon c_\mre \right) &= - \gamma_\mre \pdv{N_{\mre}}{x} + \mathcal{C}_\mre J, \label{eq:DFN+SR_ND_resc_g}\\
\pdv{i_{\mre}}{x} &= J,
\end{align}
with
\begin{align}
N_{\mre} &= -D_\mre \mathcal{B}(x) \pdv{c_{\mre}}{x} + t^+ \frac{\mathcal{C}_\mre}{\gamma_\mre} i_\mre,\\
i_{\mre} &= -\Sigma_\mre \sigma_{\mre} \mathcal{B}(x) \left(\pdv{\phi_{\mre}}{x} - 2(1-t^+) \left(1 + \pdv{f_\pm}{c_\mre} \right) \pdv{\log c_{\mre}}{x} \right). \label{eq:DFN+SR_ND_resc_j}
\end{align}
\end{subequations}

The intercalation reaction between the electrode and the electrolyte is given by
\begin{subequations}\label{eq:DFN+SR_ND_resc_reaction}
\begin{align}
J &= \begin{cases}
J_\mrn = J_{\mrn,\mathrm{int}} + J_\SR, & \text{ if } 0 \leq x \leq \ell_\mrn,\\
0, & \text{ if } \ell_\mrn < x \leq 1 - \ell_\mrp,\\
J_\mrp = J_{\mrp,\mathrm{int}}, & \text{ if } 1 - \ell_\mrp < x \leq 1,
\end{cases}
\end{align}
where the intercalation reaction kinetics are driven by
\begin{align}
J_{k,\mathrm{int}} &= \frac{\gamma_k}{\mathcal{C}_{\mathrm{r},k}} \left. \sqrt{c_{\mre} c_{k} \left(1 - c_{k} \right)} \right|_{r = 1} \sinh \left(\frac{1}{2} \left[ \lambda \left( \phi_{k} - U_k\left( \left. c_{k} \right|_{r = 1} \right) \right) - \phi_{\mre} - \frac{J_k L_{\mathrm{f}, k}}{\Sigma_{\mathrm{f}, k}} \right] \right),
\end{align}
and the side reaction kinetics are driven by
\begin{align}
J_\SR &= - \lambda^{-1} \tilde j_\SR \exp \left(- \alpha_\SR \left[ \lambda \phi_{\mrn} - \phi_{\mre}  - \frac{J_\mrn L_{\mathrm{f},\mrn}}{\Sigma_{\mathrm{f},\mrn}} \right] \right).
\end{align}
\end{subequations}

This side reaction causes a change in the porosity that follows
\begin{equation}\label{eq:DFN+SR_ND_resc_porosity}
    \pdv{\varepsilon_\mrn}{t} = \frac{\lambda}{n_\SR \tilde \gamma_\SR} J_\SR,
\end{equation}
and the corresponding film thickness can be calculated as
\begin{equation}
    L_{\mathrm{f},k} = 1 - \frac{1}{\beta_k} (\varepsilon_k - \varepsilon_{k, \mathrm{init}}).
\end{equation}

The boundary conditions at current collector ends are
\begin{subequations}\label{eq:DFN+SR_ND_resc_BC}
\begin{align}
i_{\mrn} &= i_\mathrm{app}, & N_{\mre} &= 0, & \phi_{\mre} &= 0, & \text{ at } x &= 0,\\
i_{\mrp} &= i_\mathrm{app}, & N_{\mre} &= 0, & i_{\mre} &= 0, & \text{ at } x &= 1,
\end{align}
and at the electrode-separator interfaces are
\begin{align}
i_{\mrn} &= 0, & \text{ at } x = \ell_\mrn,\\
i_{\mrp} &= 0, & \text{ at } x = 1 - \ell_\mrp.
\end{align}
Finally, the initial condition for the electrolyte is
\begin{align}
c_{\mre} &= 1, & \text{ at } t = 0.
\end{align}
\end{subequations}

\rv{The asymptotic analysis is in the limit of large $\lambda$, which physically implies small overpotentials, and we have also assumed a weak side reaction. Therefore, in the dimensionless model presented here (compared to that in \ref{sec:DFN+SR_ND})we have rescaled some dimensionless numbers with powers of $\lambda^{-1}$, which is the small parameter for the asymptotic reduction. In particular, we have used}
\begin{align}
    \gamma_\SR &= \lambda^{-1} \tilde \gamma_\SR, & j_\SR \exp \left(\alpha_\SR \lambda U_\SR \right) &= \lambda^{-1} \tilde j_\SR.
\end{align}
\rv{This choice is driven by the phenomena we want to observe: side reaction and porosity change. Note that, from the values in Table \ref{tab:parameter_values_ND}, $\gamma_\SR$ could be $\order{1}$ (e.g. for lithium plating). However, if $\gamma_\SR = \order{1}$ then the leading order porosity does not change, so assuming $\gamma_\SR = \order{\lambda^{-1}}$ allows us to capture a broader range of scenarios. Similarly, if we took $j_\SR \exp \left(\alpha_\SR \lambda U_\SR \right) = \order{\lambda^{-2}}$ instead (typical values are around $10^{-3}$) it would mean that side reactions are too small to impact the model.}

\subsection{Asymptotic reduction}\label{sec:asymptotic_reduction}
We now proceed to perform the asymptotic reduction of the model presented in the previous section. We perform the expansion in the limit of large $\lambda$ so we define our small parameter as $\lambda^{-1} \ll 1$ (which physically means small deviations from the equilibrium potential, also known as overpotentials). Then, we expand the variables of the model in powers of $\lambda^{-1}$ using the notation
\begin{equation}
    \phi_{\mrn} = \phi_{\mrn 0} + \lambda^{-1} \phi_{\mrn 1} + \lambda^{-2} \phi_{\mrn 2} + \dots,
\end{equation}
and similarly for the other variables.

\paragraph{Leading-order electrode potential} Before starting the asymptotic expansion note that, integrating \eqref{eq:DFN+SR_ND_resc_e} and using the electrode current boundary conditions in \eqref{eq:DFN+SR_ND_resc_BC}, we can determine 
\begin{equation}\label{eq:J_k}
\begin{aligned}
    \bar J_\mrn &= \frac{1}{\ell_\mrn} \int_0^{\ell_\mrn} J_\mrn \dd x = \frac{i_\mathrm{app}}{\ell_\mrn},\\
    \bar J_\mrp &= \frac{1}{\ell_\mrp} \int_{1 - \ell_\mrp}^1 J_\mrp \dd x = -\frac{i_\mathrm{app}}{\ell_\mrp}.
\end{aligned}
\end{equation}
These averaged current densities will be useful later on in the asymptotic analysis.

We now consider the leading order term of \eqref{eq:DFN+SR_ND_resc_f}, which gives
\begin{equation}
    \pdv{\phi_{k 0}}{x} = 0,
\end{equation}
and from which we can conclude that $\phi_{k 0} = \phi_{k 0} (t)$, for both $k \in \{\mrn,\mrp\}$.

\paragraph{Interface reactions} The next step is to consider the interface reaction equations \eqref{eq:DFN+SR_ND_resc_reaction} and the porosity variation equation \eqref{eq:DFN+SR_ND_resc_porosity}. Given that we are only considering a side reaction in the negative electrode, the analysis in the positive electrode will follow identically as in \cite{BrosaPlanella2021} and thus we do not reproduce the details here. For simplicity, we rewrite \eqref{eq:DFN+SR_ND_resc_reaction} using the following notation
\begin{multline}\label{eq:J_n}
    J_\mrn = J_{\mrn,\mathrm{int}} + J_\SR = j_\mrn \sinh \left( \frac{1}{2} \left[\lambda (\phi_\mrn - U_\mrn(\left. c_{\mrn} \right|_{r = 1}) ) - \phi_\mre - \frac{J_\mrn L_{\mathrm{f},\mrn}}{\Sigma_{\mathrm{f},\mrn}} \right] \right) \\ - \lambda^{-1} \tilde j_\SR \exp \left( - \alpha_\SR \left[\lambda \phi_\mrn - \phi_\mre - \frac{J_\mrn L_{\mathrm{f},\mrn}}{\Sigma_{\mathrm{f},\mrn}} \right] \right),
\end{multline}
with
\begin{align}
    j_\mrn &= \frac{\gamma_\mrn}{\mathcal{C}_{\mathrm{r},\mrn}} \left. \sqrt{c_{\mre} c_{\mrn} \left(1 - c_{\mrn} \right)} \right|_{r = 1}.
\end{align}

We have that $J_\mrn = \order{1}$, \rv{and from \eqref{eq:DFN+SR_ND_resc_porosity} we deduce that $J_\SR = \order{\lambda^{-1}}$ given that $\pdv{\varepsilon_\mrn}{t} = \order{1}$.} Then, we require that $J_{\mrn,\mathrm{int}} = \order{1}$. Expanding $J_{\mrn,\mathrm{int}}$ in powers of $\lambda^{-1}$ we have
\begin{multline}
    J_{\mrn,\mathrm{int}} = \left(j_{\mrn 0} + \lambda^{-1} j_{\mrn 1} + \dots \right) \\
    \times \sinh \left(\frac{1}{2} \left[ \lambda \left(\phi_{\mrn 0} - U_\mrn (\left. c_{\mrn 0} \right|_{r=1}) \right) + \phi_{\mrn 1} - \phi_{\mre 0} - U'_\mrn (\left. c_{\mrn 0} \right|_{r=1}) \left. c_{\mrn 1} \right|_{r=1} - \frac{J_{\mrn 0} L_{\mathrm{f},\mrn 0}}{\Sigma_{\mathrm{f},\mrn}} + \dots  \right] \right),
\end{multline}
so the large terms in the $\sinh$ need to cancel. This requires $\phi_{\mrn 0}(t) = U_\mrn (\left. c_{\mrn 0} \right|_{r=1})$. Because we know that $\phi_{\mrn 0 }$ does not depend on $x$, we can conclude that $\left. c_{\mrn 0} \right|_{r=1}$ does not depend on $x$ either. This means that the surface concentration will be identical across the particles in the negative electrode and thus, if the particles start with identical concentrations, then they will evolve with identical concentrations. From here we can deduce that the boundary flux of all the particles must be identical as well, so
\begin{equation}
    J_{\mrn,\mathrm{int} 0} = j_{\mrn 0} \sinh \left(\frac{1}{2} \left( \phi_{\mrn 1} - \phi_{\mre 0} - U'_\mrn (\left. c_{\mrn 0} \right|_{r=1}) \left. c_{\mrn 1} \right|_{r=1} - \frac{J_{\mrn 0} L_{\mathrm{f},\mrn 0}}{\Sigma_{\mathrm{f},\mrn}} \right) \right),
\end{equation}
does not depend on $x$.

Now we focus our attention on $J_\SR$, and expanding the potentials again in powers of $\lambda^{-1}$ we have
\begin{equation}
    J_\SR = - \lambda^{-1} \tilde j_\SR \exp \left( - \alpha_\SR \lambda \phi_{\mrn 0} \right) \exp \left( - \alpha_\SR \left[ \phi_{\mrn 1} - \phi_{\mre 0} - \frac{J_{\mrn 0} L_{\mathrm{f},\mrn 0}}{\Sigma_{\mathrm{f},\mrn}} + \dots  \right] \right).
\end{equation}
We can expand the rescaled exchange current density as
\begin{equation}
    \tilde j_\SR = \tilde j_{\SR 0} + \lambda^{-1} \tilde j_{\SR 1} + \dots,
\end{equation}
which yields
\begin{equation}
    J_\SR = \lambda^{-1} J_{\SR 0} = - \lambda^{-1} \tilde j_{\SR 0} \exp \left( - \alpha_\SR \lambda \phi_{\mrn 0} \right) \exp \left( - \alpha_\SR \left( \phi_{\mrn 1} - \phi_{\mre 0} - \frac{J_{\mrn 0} L_{\mathrm{f},\mrn 0}}{\Sigma_{\mathrm{f},\mrn}} \right) \right) + \dots.
\end{equation}

Now we consider the total interfacial reaction current. At leading order, from \eqref{eq:J_n} we obtain
\begin{equation}\label{eq:J_k0}
    J_{\mrn 0} = J_{\mrn,\mathrm{int} 0} = j_{\mrn 0} \sinh \left(\frac{1}{2} \left( \phi_{\mrn 1} - \phi_{\mre 0} - U'_\mrn (\left. c_{\mrn 0} \right|_{r=1}) \left. c_{\mrn 1} \right|_{r=1} - \frac{J_{\mrn 0} L_{\mathrm{f},\mrn 0}}{\Sigma_{\mathrm{f},\mrn}} \right) \right).
\end{equation}
This implies that $J_{\mrn 0}$ is homogeneous in space too and hence it must be equal to its averaged value, calculated in \eqref{eq:J_k}, so
\begin{equation}\label{eq:J_n0}
    J_{\mrn 0} = J_{\mrn,\mathrm{int} 0} = \frac{i_\mathrm{app}}{\ell_\mrn}.
\end{equation}
For the positive electrode, the analysis follows identically to that in \cite{BrosaPlanella2021}, and we find that $J_{\mrp 0} = - \frac{i_\mathrm{app}}{\ell_\mrp}$.

\paragraph{Porosity variation} For the porosity \eqref{eq:DFN+SR_ND_resc_porosity} we find that, at leading order, it is described by
\begin{equation}
    \pdv{\varepsilon_{\mrn 0}}{t} = \frac{J_{\SR 0}}{n_\SR \tilde \gamma_\SR}.
\end{equation}
This has an effect on the film thickness that appears in the reaction overpotentials. At leading order, the film thickness is calculated as
\begin{equation}
    L_{\mathrm{f},k 0} = 1 - \frac{1}{\beta_k} (\varepsilon_{k 0} - \varepsilon_{k, \mathrm{init}}).
\end{equation}

\paragraph{Electrode particles} Let's now consider the equations for electrode particles. In the positive electrode particle there is no side reaction and the analysis follows identically to that in \cite{BrosaPlanella2021}, so here we focus our attention on the negative electrode particle. At $\order{1}$ we have
\begin{equation}
\begin{aligned}
    \mathcal{C}_\mrn \pdv{c_{\mrn 0}}{t} &= \frac{1}{r^2} \pdv{}{r} \left(r^2 D_\mrn(c_{\mrn 0}) \pdv{c_{\mrn 0}}{r} \right), & \quad \text{ in } 0 < r < 1,\\
    \pdv{c_{\mrn 0}}{r} &= 0, & \quad \text{ at } r = 0,\\
    - D_\mrn(c_{\mrn 0}) \pdv{c_{\mrn 0}}{r} &= \frac{\mathcal{C}_\mrn}{\alpha_\mrn \gamma_\mrn} J_{\mrn, \mathrm{int} 0}, & \quad \text{ at } r = 1,\\
    c_{\mrn 0} &= \mu_\mrn, & \quad \text{ at } t = 0,
\end{aligned}
\end{equation}
which we can solve numerically using the expression for $J_{\mrn, \mathrm{int} 0}$ in \eqref{eq:J_n0}. In fact, because $c_{\mrn 0}$ does not depend on $x$ we can interchangeably consider the $x$-averaged concentration $\bar c_{\mrn 0}$ given that $\bar c_{\mrn 0} = c_{\mrn 0}$. At $\order{\lambda^{-1}}$ we have
\begin{equation}\label{eq:diffusion_cn1}
\begin{aligned}
    \mathcal{C}_\mrn \pdv{c_{\mrn 1}}{t} - \frac{1}{r^2} \pdv{}{r} \left(r^2 \left( D_\mrn(c_{\mrn 0}) \pdv{c_{\mrn 1}}{r} + D'_\mrn(c_{\mrn 0}) c_{\mrn 1} \pdv{c_{\mrn 0}}{r} \right) \right) &= 0, & \quad \text{ in } 0 < r < 1,\\
    - \left( D_\mrn(c_{\mrn 0}) \pdv{c_{\mrn 1}}{r} + D'_\mrn(c_{\mrn 0}) c_{\mrn 1} \pdv{c_{\mrn 0}}{r} \right) &= 0, & \quad \text{ at } r = 0,\\
    - \left( D_\mrn(c_{\mrn 0}) \pdv{c_{\mrn 1}}{r} + D'_\mrn(c_{\mrn 0}) c_{\mrn 1} \pdv{c_{\mrn 0}}{r} \right) &= \frac{\mathcal{C}_\mrn}{\alpha_\mrn \gamma_\mrn} J_{\mrn, \mathrm{int} 1}, & \quad \text{ at } r = 1,\\
    c_{\mrn 1} &= 0, & \quad \text{ at } t = 0.
\end{aligned}
\end{equation}
We can now $x$-average \eqref{eq:diffusion_cn1} to obtain an equation for $\bar c_{\mrn 1}$. We also make use of the fact that $\bar c_{\mrn 0} = c_{\mrn 0}$ to simplify the resulting equation into
\begin{equation}\label{eq:diffusion_cn1_bar}
\begin{aligned}
    \mathcal{C}_\mrn \pdv{\bar c_{\mrn 1}}{t} - \frac{1}{r^2} \pdv{}{r} \left(r^2 \left( D_\mrn(\bar c_{\mrn 0}) \pdv{\bar c_{\mrn 1}}{r} + D'_\mrn(\bar c_{\mrn 0}) \bar c_{\mrn 1} \pdv{\bar c_{\mrn 0}}{r} \right) \right) &= 0, & \quad \text{ in } 0 < r < 1,\\
    - \left( D_\mrn(\bar c_{\mrn 0}) \pdv{\bar c_{\mrn 1}}{r} + D'_\mrn(\bar c_{\mrn 0}) \bar c_{\mrn 1} \pdv{\bar c_{\mrn 0}}{r} \right) &= 0, & \quad \text{ at } r = 0,\\
    - \left( D_\mrn(\bar c_{\mrn 0}) \pdv{\bar c_{\mrn 1}}{r} + D'_\mrn(\bar c_{\mrn 0}) \bar c_{\mrn 1} \pdv{\bar c_{\mrn 0}}{r} \right) &= \frac{\mathcal{C}_\mrn}{\alpha_\mrn \gamma_\mrn} \bar J_{\mrn, \mathrm{int 1}}, & \quad \text{ at } r = 1,\\
    \bar c_{\mrn 1} &= 0, & \quad \text{ at } t = 0.
\end{aligned}
\end{equation}

We need to calculate $\bar J_{\mrn, \mathrm{int 1}}$. Averaging the $\order{\lambda^{-1}}$ term of \eqref{eq:J_n} over the negative electrode yields
\begin{equation}
    \bar J_{\mrn 1} = \bar J_{\mrn, \mathrm{int} 1} + \bar J_{\SR 0},
\end{equation}
and from \eqref{eq:J_k} we deduce that $\bar J_{\mrn 1} = 0$, so we conclude that $\bar J_{\mrn, \mathrm{int} 1} = - \bar J_{\SR 0}$. Therefore, for the negative electrode particle we cannot conclude that $\bar c_{\mrn 1} = 0$ as it occurs with the positive electrode particle (see \cite{BrosaPlanella2021}). Instead, $\bar c_{\mrn 1}$ captures the contribution of the side reaction. To determine it we need to solve \eqref{eq:diffusion_cn1_bar}, which can be easily done numerically. \rv{Note that $x$-averaging the equations over each electrode is a key step to simplify the equations. This step and its consequences are discussed in detail in \cite{Marquis2019}.}

\paragraph{Electrolyte} We now consider the equations for the electrolyte \eqref{eq:DFN+SR_ND_resc_g}-\eqref{eq:DFN+SR_ND_resc_j}. First we calculate the leading order current, which is determined by 
\begin{align}
    \pdv{i_{\mre 0}}{x} &= J_0 = \begin{cases}
    J_{\mrn 0} = \frac{i_\mathrm{app}}{\ell_\mrn}, & \text{ if } 0 \leq x < \ell_\mrn,\\
    0, & \text{ if } \ell_\mrn \leq x < 1 - \ell_\mrp,\\
    J_{\mrp 0} = -\frac{i_\mathrm{app}}{\ell_\mrp}, & \text{ if } 1 - \ell_\mrp \leq x \leq 1.
    \end{cases}
\end{align}
Integrating and imposing continuity of current we find
\begin{align}\label{eq:i_e0}
    i_{\mre 0} &= \begin{cases}
    \frac{i_\mathrm{app}}{\ell_\mrn}x, & \text{ if } 0 \leq x < \ell_\mrn,\\
    i_\mathrm{app}, & \text{ if } \ell_\mrn \leq x < 1 - \ell_\mrp,\\
    \frac{i_\mathrm{app}}{\ell_\mrp} (1 - x), & \text{ if } 1 - \ell_\mrp \leq x \leq 1.
    \end{cases}
\end{align}

Now we consider ion concentration in the electrolyte. At leading order it is governed by
\begin{equation}
\begin{aligned}
    \mathcal{C}_\mre \gamma_\mre \pdv{}{t} \left( \varepsilon_0 c_{\mre 0} \right) &=  \pdv{}{x} \left(\gamma_\mre D_\mre (c_{\mre 0}) \mathcal{B}_0(x) \pdv{c_{\mre 0}}{x} + (1 - t^+(c_{\mre 0})) \mathcal{C}_\mre i_{\mre 0} \right), & \text{ in } 0 \leq x \leq 1,\\
    \pdv{c_{\mre 0}}{x} &= 0, & \text{ at } x = 0,1,\\
    c_{\mre 0} &= 1, & \text{ at } t = 0,
\end{aligned}
\end{equation}
where $i_{\mre 0}$ is piecewise linear as determined in \eqref{eq:i_e0}. This equation can be solved numerically to compute $c_{\mre 0}$, as it is done in the standard SPMe.

Finally, we need to calculate the electrolyte potential. At leading order we have
\begin{align}
    i_{\mre 0} &= -\Sigma_\mre \sigma_{\mre} (c_{\mre 0}) \mathcal{B}_0(x) \left(\pdv{\phi_{\mre 0}}{x} - 2(1-t^+(c_{\mre 0})) \left(1 + \pdv{f_\pm}{c_{\mre 0}} \right) \pdv{\log c_{\mre 0}}{x} \right),
\end{align}
which we can rearrange into
\begin{align}
    \pdv{\phi_{\mre 0}}{x} &= -\frac{i_{\mre 0}}{\Sigma_\mre \sigma_{\mre 0} \mathcal{B}_0(x)} + 2 (1 - t^+(c_{\mre 0})) \left(1 + \pdv{f_\pm}{c_{\mre 0}} \right) \pdv{\log c_{\mre 0}}{x}.
\end{align}
Integrating from $0$ to $x$ and using the fact that $\phi_{\mre 0} = 0$ at $x = 0$ \eqref{eq:DFN+SR_ND_resc_BC} we obtain
\begin{align}
    \phi_{\mre 0}  &= - \int_0^x \frac{i_{\mre 0}}{\Sigma_\mre \sigma_\mre (c_{\mre 0}(s, t)) \mathcal{B}_0(s)} \dd s + \int_0^x 2 (1 - t^+(c_{\mre 0}(s, t))) \left(1 + \pdv{f_\pm}{c_{\mre 0}} \right) \pdv{\log c_{\mre 0}(s,t)}{s} \dd s.
\end{align}

\paragraph{Higher-order electrode potential} We finally calculate the higher order terms for the electrode potentials, which will allow us to determine the terminal voltage. At $\order{\lambda^{-1}}$ we have
\begin{align}
    -\Sigma_{k}\pdv[2]{\phi_{k 1}}{x} = - J_{k 0},
\end{align}
for $k \in \{\mrn,\mrp\}$. Integrating, we find
\begin{align}
    \phi_{k 1} &= \frac{J_{k 0}}{2 \Sigma_k} x^2 + A_k x + B_k,
\end{align}
where $A_k$ and $B_k$ are integration constants. We can determine $A_k$ \rv{from the boundary conditions for $i_k$ \eqref{eq:DFN+SR_ND_resc_BC},} obtaining
\begin{equation}
\begin{aligned}
    \phi_{\mrn 1} &= - \frac{i_\mathrm{app} (2 \ell_\mrn - x) x}{2 \ell_\mrn \Sigma_\mrn} + B_\mrn,\\
    \phi_{\mrp 1} &= \frac{i_\mathrm{app} (2 (1 - \ell_\mrp) - x) x}{2 \ell_\mrp \Sigma_\mrp} + B_\mrp.
\end{aligned}
\end{equation}
We calculate $B_k$ from the interface reaction as it relates the electrode and electrolyte potentials. From \eqref{eq:J_k0} we have
\begin{align}\label{eq:inverted_BV}
    \phi_{k 1} - \phi_{\mre 0} - \left. U'_k(c_{k 0}) c_{k 1} \right|_{r = 1} - \frac{J_{k 0} L_{\mathrm{f}, k 0}}{\Sigma_{\mathrm{f}, k}} = 2 \arcsinh \left( \frac{J_{k 0}}{j_{k 0}} \right).
\end{align}
We now average \eqref{eq:inverted_BV} over the corresponding electrode. This will not affect $B_k$, as they are constants, and will make the calculations simpler.

We now need to distinguish between the positive and the negative electrode. For the positive electrode we have $\bar c_{\mrp 1} = 0$ (see details in \cite{BrosaPlanella2021}), while for the negative electrode this is not true due to the side reaction. Then, we have
\begin{equation}
\begin{aligned}
    B_\mrn &= \frac{i_\mathrm{app} \ell_\mrn}{3 \Sigma_\mrn} + \left. U'_\mrn(\bar c_{\mrn 0}) \bar c_{\mrn 1} \right|_{r = 1} + \frac{i_\mathrm{app} \bar L_{\mathrm{f}, \mrn 0}}{\ell_\mrn \Sigma_{\mathrm{f},\mrn}} - \frac{1}{\ell_\mrn \Sigma_\mre} \int_0^{\ell_\mrn} \int_0^x \frac{i_{\mre 0}(s,t) \dd s}{\sigma_\mre (c_{\mre 0}(s,t)) \mathcal{B}_0(s)} \dd x \\
    & \quad + \frac{1}{\ell_\mrn} \int_0^{\ell_\mrn} \int_0^x 2 (1 - t^+(c_{\mre 0}(s, t))) \left(1 + \pdv{f_\pm}{c_{\mre 0}} \right) \pdv{\log c_{\mre 0}(s,t)}{s} \dd s \dd x + \frac{2}{\ell_\mrn} \int_0^{\ell_\mrn} \arcsinh\left(\frac{i_\mathrm{app}}{\ell_\mrn j_{\mrn 0}} \right) \dd x,\\
    B_\mrp &= - \frac{i_\mathrm{app}(2 \ell_\mrp^2 - 6 \ell_\mrp + 3)}{6 \ell_\mrp \Sigma_\mrp} - \frac{i_\mathrm{app} \bar L_{\mathrm{f}, \mrp 0}}{\ell_\mrp \Sigma_{\mathrm{f},\mrp}} - \frac{1}{\ell_\mrp \Sigma_\mre} \int_{1 - \ell_\mrp}^1 \int_0^x \frac{i_{\mre 0}(s,t) \dd s}{\sigma_\mre (c_{\mre 0}(s,t)) \mathcal{B}_0(s)} \dd x \\
    & \quad + \frac{1}{\ell_\mrp} \int_{1 - \ell_\mrp}^1 \int_0^x 2 (1 - t^+(c_{\mre 0}(s, t))) \left(1 + \pdv{f_\pm}{c_{\mre 0}} \right) \pdv{\log c_{\mre 0}(s,t)}{s} \dd s \dd x - \frac{2}{\ell_\mrp} \int_{1 - \ell_\mrp}^1 \arcsinh\left(\frac{i_\mathrm{app}}{\ell_\mrp j_{\mrp 0}} \right) \dd x.
\end{aligned}
\end{equation}


\rv{With the potential expressions determined at $\order{\lambda^{-1}}$, we have all the components we need for the SPMe+SR model. We have leading order expressions for each variable, except for the particle concentrations and electrode potentials for which we have also computed the first order corrections. Then, we define each variable as: 
\begin{equation}
\begin{aligned}
    \bar c_k &\approx \bar c_{k 0} + \lambda^{-1} \bar c_{k 1},\\
    c_\mre &\approx c_{\mre 0},\\
    \varepsilon &\approx \varepsilon_0,\\
    L_{\mathrm{f},k} &\approx L_{\mathrm{f},k 0},\\
    \phi_{k} &\approx \phi_{k 0} + \lambda^{-1} \phi_{k 1},\\
    \phi_{\mre} &\approx \phi_{\mre 0}.
\end{aligned}
\end{equation}
The full expressions for the dimensionless SPMe+SR are provided in \ref{sec:model_summary}.}

\section{Results \& discussion}\label{sec:results}



After introducing the SPMe+SR model in Section \ref{sec:SPMe+SR} and providing its formal derivation from the DFN+SR model using asymptotic methods in Section \ref{sec:derivation_SPMe_SR}, we now validate it by comparing its performance against the full DFN+SR model. As detailed in this section, we find that the SPMe+SR provides very similar results to the DFN+SR for the three tested scenarios (SEI growth, lithium plating and both effects combined), while being an order of magnitude smaller and significantly faster. Both models have been implemented in PyBaMM \rv{(v22.10)}, an open-source battery modelling package \cite{Sulzer2021}. The code to reproduce the results of this article is publicly available (see ``Data and Code availability'') In order to solve both models, we have used the method of lines \cite{Schiesser2016} with a finite volume method for the spatial discretisation \cite{LeVeque2002}, \rv{which ensures lithium conservation}. For the results here, we use 20 points in each particle and in each electrode and separator. For the solvers, we have tested \texttt{scikits.odes} \cite{Malengier2018} and CasADI \cite{Andersson2019}, both readily available in PyBaMM. The parameters used in the simulations are shown in Tables \ref{tab:parameter_values_LGM50} and \ref{tab:parameter_values_SR}.

\begin{table}
\centering
\begin{tabular}{| c c l c c c |}
\hline
\textbf{Symbol} & \textbf{Units} & \textbf{Description} & \textbf{Pos.} & \textbf{Sep.} & \textbf{Neg.} \\ \hline
$L_{k}$ & m & Thickness & $75.6\E{-6}$ & $12\E{-6}$ & $85.2\E{-6}$ \\
$R_{k}$ & m & Radius of electrode particles & $5.22\E{-6}$ & - & $5.86\E{-6}$ \\
$a_{k}$ & m$^{-1}$ & Particle surface area density & $3.82\E{5}$ & - & $3.84\E{5}$ \\
$D_{k}$ & $\mathrm{m}^2 \; \mathrm{s}^{-1}$ & Lithium diffusivity in particles & $4\E{-15}$ & - & $3.3\E{-14}$ \\
$\sigma_{k}$ & $\mathrm{S} \; \mathrm{m}^{-1}$ & Electrode conductivity & 0.18 & - & 215 \\
$c_{k, \mathrm{init}}$ & $\mathrm{mol} \; \mathrm{m}^{-3}$ & Initial particle concentration & 17038 & - & 29866 \\
$c_{k}^{\max}$ & $\mathrm{mol} \; \mathrm{m}^{-3}$ & Max. particle concentration & 63104 & - & 33133 \\
$U_k$ & V & Open-circuit potential & \eqref{eq:U_p} & - & \eqref{eq:U_n} \\
$m_k$ & $\mathrm{A} \; \mathrm{m}^{-2} \left(\mathrm{mol} \; \mathrm{m}^{-3}\right)^{-1.5}$ & Reaction rate & $3.42\E{-6}$ & - & $6.48\E{-7}$ \\ \hline
$\varepsilon_{k}$ & - & Electrolyte volume fraction & 0.335 & 0.47 & 0.25 \\
$D_{\mre}$ & $\mathrm{m}^2 \; \mathrm{s}^{-1}$ & Electrolyte diffusivity & \multicolumn{3}{c |}{\eqref{eq:D_e}} \\
$\sigma_{\mre}$ & $\mathrm{S} \; \mathrm{m}^{-1}$ & Electrolyte conductivity & \multicolumn{3}{c |}{\eqref{eq:sigma_e}} \\
$t^+$ & - & Transfer number & \multicolumn{3}{c |}{0.2594} \\
$c_{\mre,\mathrm{init}}$ & $\mathrm{mol} \; \mathrm{m}^{-3}$ & Initial electrolyte concentration & \multicolumn{3}{c |}{1000} \\ \hline
$i_{\mathrm{app}}$ & $\mathrm{A} \; \mathrm{m}^{-2}$ & Applied current density & \multicolumn{3}{c |}{$48.69 C$}\\
$F$ & $\mathrm{C} \; \mathrm{mol}^{-1}$ & Faraday constant & \multicolumn{3}{c |}{96485}\\
$R$ & $\mathrm{J} \; \mathrm{K}^{-1} \; \mathrm{mol}^{-1}$ & Gas constant & \multicolumn{3}{c |}{8.314}\\
$T$ & K & Reference temperature & \multicolumn{3}{c |}{298}\\ \hline
\end{tabular}
\caption{Dimensional parameters for the electrochemical model, corresponding to the LG M50 cell and taken from \cite{Chen2020}. The $C$ in the definition of $i_\mathrm{app}$ corresponds to the C-rate of the experiment. The model needs additional parameters for the corresponding degradation model, which are provided in Table \ref{tab:parameter_values_SR}.}
\label{tab:parameter_values_LGM50}
\end{table}

\begin{table}
\centering
\begin{tabular}{| c c l c c|}
\hline
\textbf{Symbol} & \textbf{Units} & \textbf{Description} & \textbf{Value} & \textbf{Ref.}\\ \hline
$k_\SEI$ & $\mathrm{m} \; \mathrm{s}^{-1}$ & Reaction rate of SEI growth & $1\E{-12}$ & \cite{Yang2017} \\
$c_{\SEI,\mathrm{init}}$ & $\mathrm{mol} \; \mathrm{m}^{-3}$ & SEI material concentration in electrolyte & $4541$ & \cite{Ploehn2004} \\
$U_\SEI$ & V & Open-circuit potential of SEI reaction & 0 & \cite{Safari2009,Yang2017}* \\
$D_\SEI$ & $\mathrm{m}^2 \; \mathrm{s}^{-1}$ & Diffusivity of the SEI material & $2\E{-19}$ & adj. \\
$M_\SEI$ & $\mathrm{kg} \; \mathrm{mol}^{-1}$ & Molar weight of SEI material & $0.162$ & \cite{Safari2009}\\
$\rho_\SEI$ & $\mathrm{kg} \; \mathrm{m}^{-3}$ & Density of SEI material & $1690$ & \cite{Borodin2006}\\
$n_\SEI$ & - & Number of electrons in SEI reaction & 2 & \cite{Safari2009} \\
$\sigma_\SEI$ & $\mathrm{S} \; \mathrm{m}^{-1}$ & Conductivity of the SEI layer & $5\E{-6}$ & \cite{Safari2009} \\
$L_{\mathrm{f},\mathrm{init}}$ & $\mathrm{m}$ & Initial thickness of the SEI layer & $5\E{-9}$ & \cite{Safari2009} \\ \hline
$k_\Li$ & $\mathrm{m} \; \mathrm{s}^{-1}$ & Reaction rate of plating reaction & $1\E{-11}$ & calc. \\
$U_\Li$ & V & Open-circuit potential of plating reaction & 0 & \cite{Yang2017} \\
$M_\Li$ & $\mathrm{kg} \; \mathrm{mol}^{-1}$ & Molar weight of lithium & $6.94\E{-3}$ & \cite{Haynes2014} \\
$\rho_\Li$ & $\mathrm{kg} \; \mathrm{m}^{-3}$ & Density of lithium & 534 & \cite{Haynes2014} \\
$n_\Li$ & - & Number of electrons in plating reaction & 1 & \cite{Yang2017} \\ 
$L_{\mathrm{f},\mathrm{init}}$ & $\mathrm{m}$ & Initial thickness of the plated lithium & $0$ & assum. \\ \hline
\end{tabular}
\caption{Dimensional parameters for the side reaction models. The parameters have mostly been taken to match \cite{Yang2017} but, where applicable, we have cited the original source. The open-circuit potential of the SEI reaction $U_\SEI$ is not provided in \cite{Yang2017}, but we set it to zero as its effect has been absorbed by $k_\SEI$ (see \cite{Safari2009} for a detailed discussion). The diffusivity in the SEI material $D_\SEI$ has been adjusted to yield reasonable results for the LG M50. The reaction rate of lithium plating $k_\Li$ has been calculated to match the value provided in \cite{Yang2017}, and the initial thickness of plated lithium has been assumed to be zero. Finally, plated lithium has been assumed to be a perfect conductor, so the conductivity is not needed.}
\label{tab:parameter_values_SR}
\end{table}

\begin{figure}
    \centering
    \begin{subfigure}[c]{0.49\textwidth}
        \centering
        \includegraphics[scale=1]{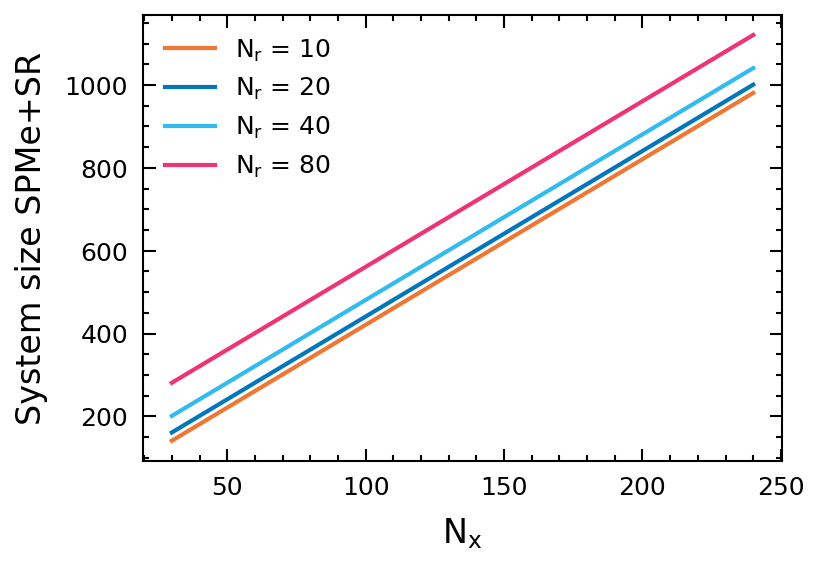}
        \caption{SPMe+SR}
    \end{subfigure}\hfill
    \begin{subfigure}[c]{0.49\textwidth}
        \centering
        \includegraphics[scale=1]{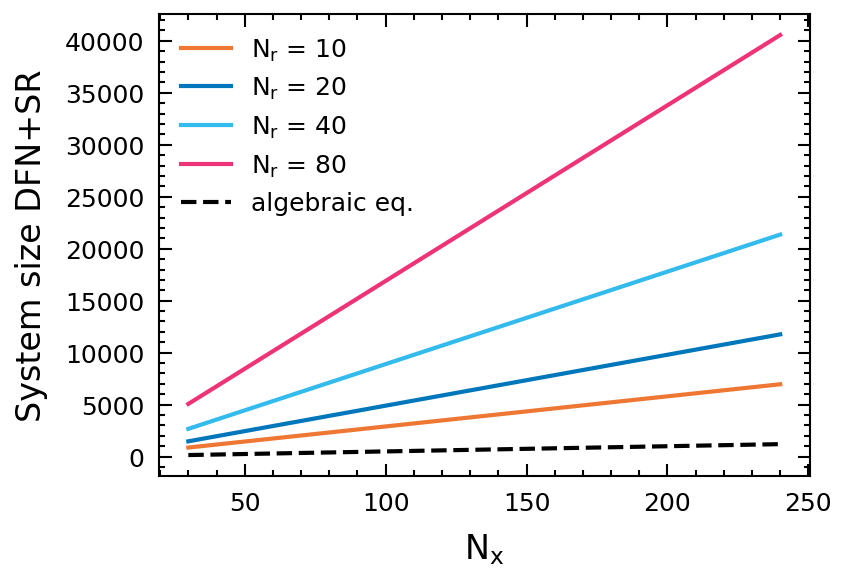}
        \caption{DFN+SR}
    \end{subfigure}
\caption{Comparison of the number of differential (colour solid lines) and algebraic equations (black dashed line) as a function of the mesh size in the electrodes and separator ($N_x$) and in the particles ($N_r$). Here we have assumed only one side reaction ($k=1$).}
\label{fig:system_size}
\end{figure}

Before showing the simulation results, we compare the size of the systems of equations obtained after discretisation for both SPMe+SR and DFN+SR. Assuming that the discretisation has $N_x$ points in each electrode and separator and $N_r$ points in each particle, and defining $k$ as the number of side reactions, the sizes of the systems follow the expressions shown in Table \ref{tab:system_size} and plotted in Figure~\ref{fig:system_size}. We observe two main differences on how the systems scale. First, the number of differential equations in SPMe+SR scales linearly with each variable, while in DFN+SR it scales as $N_x N_r$. Second, the number of algebraic equations in the SPMe+SR is zero (if the current is prescribed) while for the DFN+SR it scales linearly with $N_x$. In the particular case of $N_x = N_r = 20$ and $k = 1$, which is the one used in the results here, the SPMe+SR consists only of 121 differential equations (and no algebraic equations), while the DFN+SR consists of 881 differential equations and 100 algebraic equations. \rv{Note that this applies only to the case where current is prescribed. If instead voltage or power are prescribed, we need to introduce an additional algebraic equation for both models. This means that, in that case, the SPMe+SR consists of a system of DAEs, with a single algebraic equation.}

\begin{table}
\centering
\begin{tabular}{| l | c | c |}
\cline{2-3}
\multicolumn{1}{l |}{} & \textbf{\# of differential equations} & \textbf{\# of algebraic equations} \\ \hline
\textbf{SPMe+SR} & $2 N_r + (3 + k) N_x + 1$ & 0 \\ \hline
\textbf{DFN+SR} & $2 N_x N_r + (3 + k) N_x + 1$ & $5 N_x$ \\ \hline
\end{tabular}
\caption{Size of discretised systems for the SPMe+SR and DFN+SR. $N_x$ is the number of points in each electrode and separator, $N_r$ is the number of points in each particle, and $k$ is the number of side reactions. \rv{This applies only to the case where the applied current is prescribed, otherwise an additional algebraic equation is introduced.}}
\label{tab:system_size}
\end{table}

In addition to the size of the system, we study the solving time for different solvers, experiments and mesh sizes. In practical applications we might be constrained by the simulation platform on the solvers to use, so here we test two different families of solvers to see the variability: the CasADI solvers and the \texttt{scikits.odes} solvers. Both solvers are built on top of the suite SUNDIALS \cite{Hindmarsh2005}, and can handle both ODEs (using SUNDIALS' CVODE) and DAEs (using SUNDIALS' IDA). We also consider two different experiments: a single constant current (CC) discharge, and 10 cycles of CC discharge and constant current constant voltage (CCCV) charge. In all cases, the discharge rate is 1C and the charge rate is C/2, followed by a 4.2 V CV step with a C/20 cut-off current. Note that in the CC discharge for the SPMe+SR, the resulting system consists only of ODEs so we can use the \texttt{ode} solver, while for the CCCV scenario and the DFN+SR model we need to use the \texttt{dae} solver. For each experiment and model we record the solving time for different mesh sizes. Before going ahead, we want to remark that the objective here is not to provide a thorough performance comparison. We leave the thorough comparison as an area of future work, given that it would require a much more detailed study of model parameters, mesh sizes and solver settings. The computational times have been calculated on a laptop with an \rv{11th Generation Intel Core i7-1185G7 (3.00Ghz) processor and 16 GB RAM}.

As shown in Table~\ref{tab:solving_times}, we observe a significant reduction in the SPMe+SR compared to the DFN+SR. \rv{For the CasADI solvers we observe that the SPMe+SR is between 2 and 18 times faster than the DFN+SR. These differences strongly depend on the \texttt{dt\_max} parameter of the PyBaMM CasADI solver, which controls the detection of events (e.g. cut-off voltage). To make a fair comparison we have set $\texttt{dt\_max} = 1000$, relaxing the event detection which hinders performance of the ODE solver.} For the \texttt{scikits.odes} solvers we observe a much more dramatic speed-up which is a lot more sensitive to the mesh sizes and the experiments, \rv{ranging from a factor of 40 in the CC discharge for the coarser mesh, to a factor of over 150 in the CC discharge for the finer mesh. In summary, even though the choice of solver and its settings need to be studied in a case by case basis, we find that SPMe+SR is consistently faster than the DFN+SR across different experiments and solvers.}

\begin{table}
\centering
\begin{tabular}{| l | c | c | c | c | c|}
\cline{3-6}
\multicolumn{2}{ c }{} & \multicolumn{2}{| c |}{CasADI} & \multicolumn{2}{| c |}{\texttt{scikits.odes}} \\ \cline{3-6}
\multicolumn{2}{ c |}{} & SPMe+SR & DFN+SR & SPMe+SR & DFN+SR \\ \hline
\multirow{4}{*}{\rv{CC discharge}} & $N_x = N_r = 20$ & $0.06 \pm 0.01$ & $1.03 \pm 0.36$ & $0.15 \pm 0.02$ & $5.89 \pm 0.05$ \\
& $N_x = 20$, $N_r = 40$ & $0.06 \pm 0.01$ & $1.13 \pm 0.17$ & $0.16 \pm 0.01$ & $19.42 \pm 0.50$ \\
& $N_x = 40$, $N_r = 20$ & $0.20 \pm 0.02$ & $1.74 \pm 0.43$ & $0.41 \pm 0.01$ & $27.12 \pm 0.23$ \\
& $N_x = N_r = 40$ & $0.22 \pm 0.02$ & $2.20 \pm 0.19$ & $0.45 \pm 0.01$ & $69.53 \pm 1.84$ \\ \hline
\multirow{4}{*}{\rv{CCCV cycles}} & $N_x = N_r = 20$ & $2.59 \pm 0.03$ & $12.21 \pm 0.51$ & $5.42 \pm 0.03$ & $165.58 \pm 0.60$ \\
& $N_x = 20$, $N_r = 40$ & $2.70 \pm 0.04$ & $17.10 \pm 0.62$ & $6.80 \pm 0.05$ & $524.39 \pm 11.81$ \\
& $N_x = 40$, $N_r = 20$ & $9.80 \pm 0.07$ & $22.22 \pm 0.80$ & $13.80 \pm 0.07$ & $697.54 \pm 5.95$ \\
& $N_x = N_r = 40$ & $10.07 \pm 0.09$ & $32.33 \pm 1.56$ & $16.60 \pm 0.03$ & $1994.43 \pm 12.50$ \\ \hline
\end{tabular}
\caption{Solving times \rv{(in seconds)} for the SPMe+SR and DFN+SR for different experiments, mesh sizes and solvers. The two experiments are a 1C CC discharge and 10 cycles of a 1C CC discharge followed by a C/2 4.2 V CCCV charge, with a C/20 cut-off current. The mesh sizes comprise the four different combinations of $N_x,N_r \in \{20, 40\}$. The solvers studied are the CasADI solvers \cite{Andersson2019} and the \texttt{scikits.odes} solvers \cite{Malengier2018}. \rv{The values presented are the mean and standard deviation of 10 identical and independent simulations.}}
\label{tab:solving_times}
\end{table}


We validate the SPMe+SR model for three different types of side reaction: SEI growth, lithium plating, and both reactions together. The SEI growth model is a solvent diffusion model with reaction, as defined in \eqref{eq:j_SEI}, while for the lithium plating we use the irreversible model defined in \eqref{eq:j_Li}. The combined model is defined by \eqref{eq:j_SEI}-\eqref{eq:porosity_combined_SR}. We simulate 1000 cycles, where each cycle is defined as a 1C CC discharge, followed by a C/2 and 4.2 V CCCV charge. The cut-off voltages are 4.2 V and 2.5 V, while the cut-off current for the CV step is C/20. Other combinations of charge (C/2 and C/3) and discharge (1C and 2C) rates have been considered, finding very similar results. The plots for these additional cycling conditions can be found in the Supplementary Information. 

\begin{figure}
    \centering
    \includegraphics[scale=1]{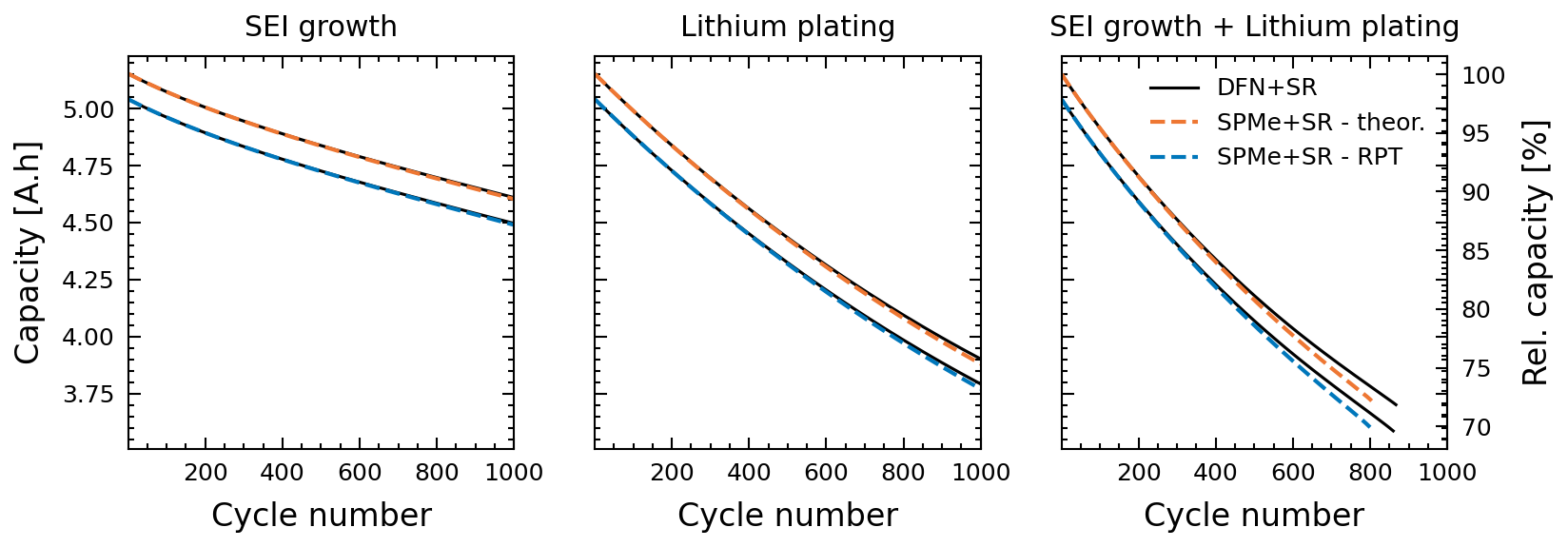}
    \caption{Comparison between the SPMe+SR and the DFN+SR of the discharge capacity as a function of the cycle number. The theoretical capacity corresponds to that of the total cyclable lithium (i.e. equivalent to an infinitely slow discharge) while the RPT capacity corresponds to a discharge rate of C/3.}
    \label{fig:results_capacity}
\end{figure}

\begin{figure}
    \centering
    \includegraphics[scale=1]{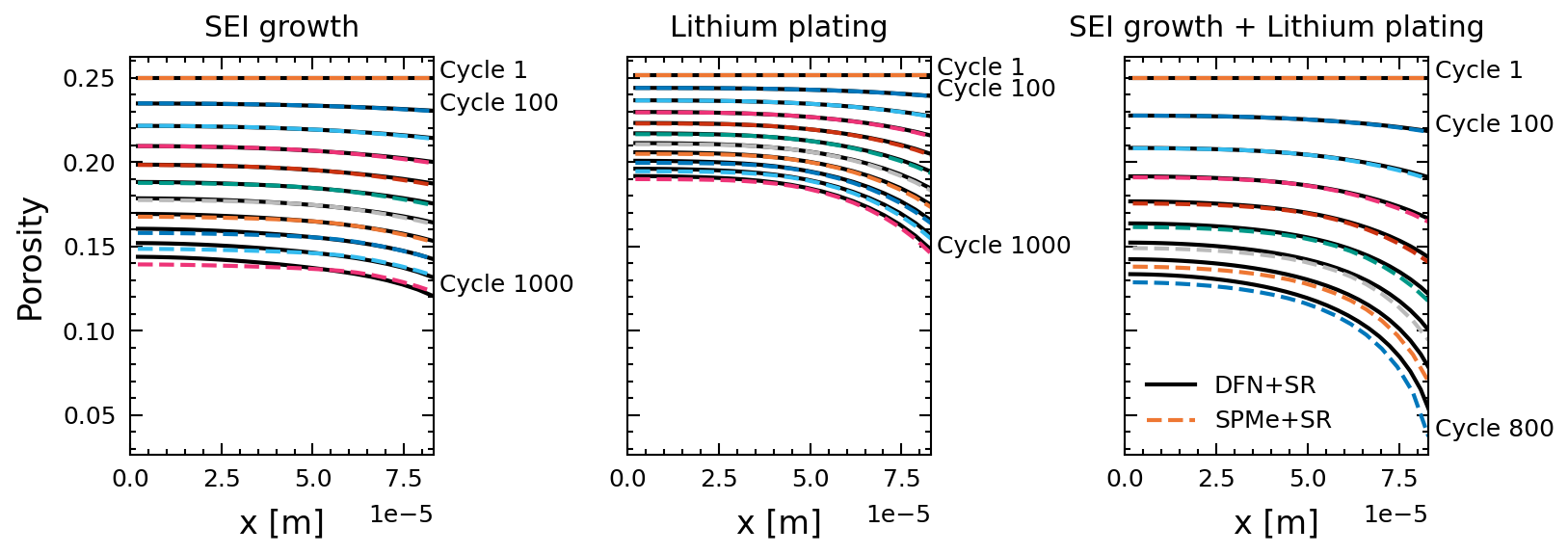}
    \caption{Comparison between the SPMe+SR and the DFN+SR of the negative electrode porosity. For each case, we plot the capacity every 100 cycles (plus the initial cycle).}
    \label{fig:results_porosity}
\end{figure}

\begin{figure}[htp]
    \centering
    \includegraphics[scale=1]{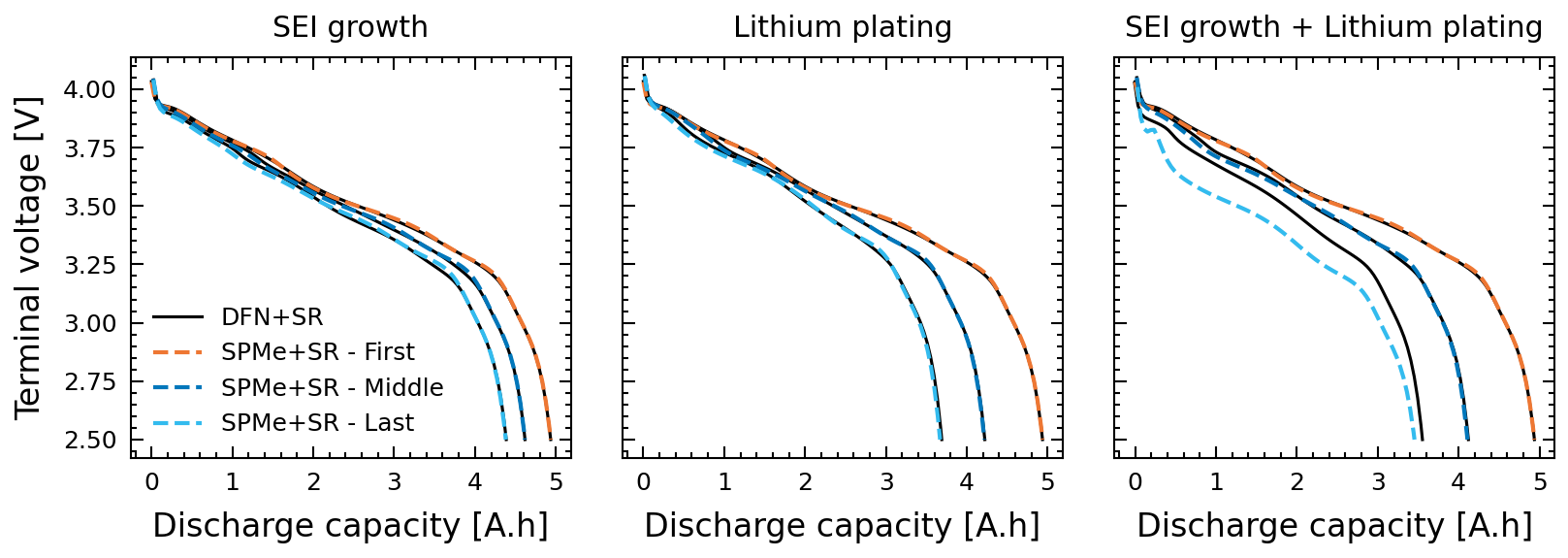}
    \caption{Comparison between the SPMe+SR and the DFN+SR of the voltage profile at different cycles. For each side reaction we plot the first, middle and last cycle. For SEI growth and lithium plating, middle cycle is 500 and final cycle is 1000, while for the SEI growth + lithium plating the middle cycle is 402 and the final cycle is 804. \rv{The cycles are different for the latter, as electrolyte depletion is hit at cycle 804 and the simulation must be stopped.}}
    \label{fig:results_voltage}
\end{figure}

For all figures comparing SPMe+SR and DFN+SR, the colour dashed lines correspond to the SPMe+SR (different colours for different scenarios), while the solid black lines represent the DFN+SR (i.e. the benchmark). We find that across the various plots, the SPMe+SR results are very similar to the DFN+SR ones. Figure~\ref{fig:results_capacity} shows both the theoretical capacity (\rv{as defined in \cite{Mohtat2019}, and} equivalent to an infinitely slow discharge) and the capacity for a C/3 Reference Performance Test (RPT). RPTs are simulated in parallel by taking the initial state at the beginning of each tested cycle (i.e. fully charged battery) and then performing a C/3 CC discharge and taking the discharge capacity at the end of discharge as the RPT capacity for that cycle. In this particular case, we perform RPT every 10 cycles (i.e. first cycle plus all multiples of 10). As expected, the capacity of the battery fades over time, and it ages faster when both SEI growth and lithium plating are included. Figure~\ref{fig:results_porosity} shows the spatially distributed porosity every 100 cycles. We observe that both side reactions tend to grow near the separator (right hand side of the plots) given that the overpotentials are larger there. Finally, Figure~\ref{fig:results_voltage} shows the voltage profiles versus discharge capacity for the first, middle and final cycles. We observe that the agreement between both models is quite good except for the final cycle of the model accounting for both SEI growth and lithium plating, which is after 804 cycles (rather than 1000) due to electrolyte depletion. This discrepancy is due to a mismatch in the electrolyte states (both concentration and potential) caused by the underestimation of the porosity by SPMe+SR (see Figure~\ref{fig:results_porosity}). \rv{Inspecting the model equations, we see that a lower porosity results in a larger reaction overpotential, which in turn leads to stronger side reactions and even lower porosity. However, further work is required to validate this hypothesis.} From the three figures, we can observe a very good agreement between the SPMe+SR and DFN+SR for a wide range of degradation models and operating conditions, with deviations only becoming noticeable with cycling for the combined SEI and lithium plating due to electrolyte states.

Overall, we find that the SPMe+SR captures accurately the various scales and features involved in the DFN+SR while being much simpler and faster to simulate. In particular, we observe that the SPMe+SR, despite being a single particle model, can predict not only the global states of the battery, such as capacity and voltage, but also spatially distributed internal states like the porosity. Moreover, this accuracy spans across the two time scales involved in the simulation: single cycle (e.g. voltage vs discharge capacity for a given cycle in Figure~\ref{fig:results_voltage}) and battery lifetime (e.g. capacity vs cycle number in Figure~\ref{fig:results_capacity}). The main discrepancies are observed only after hundreds of cycles and even then they are within a reasonable range given how \rv{simple} the reduced model is. Moreover, for practical applications, we could use control techniques to correct these deviations \cite{Li2021}. The main limitation of the SPMe+SR model can be observed in Figure~\ref{fig:results_capacity} for the SEI growth + lithium plating model, in which we notice that the simulation for the SPMe+SR stops earlier than the DFN+SR one. This is caused by electrolyte depletion during the charge step, which means that due to the lower porosity, the ions in the electrolyte intercalate at a faster rate than they can be transported in the electrolyte. This leads to a region of the negative electrode near the current collector not receiving ions. In this case, the SPMe+SR breaks down as the assumption of uniform behaviour for all electrode particles is no longer true, and it is an area for future work to explore how the SPMe+SR (or other reduced models) can be extended to account for such situations.

\section{Conclusions}



We have introduced the Single Particle Model with electrolyte and Side Reactions (SPMe+SR), a reduced physics-based model for electrochemical degradation. To the best of our knowledge, this is the first time that a model of this class has been formally derived from a Doyle-Fuller-Newman model with Side Reactions (DFN+SR), as opposed to other similar models in the literature which have been posed in an \emph{ad hoc} manner that often leads to inconsistencies. To derive the SPMe+SR we have used asymptotic methods which, by exploiting some \emph{a priori} assumptions on the size of the dimensionless quantities of the model, allow us to simplify the DFN+SR model in a generic, systematic and flexible way. The analysis is based on the assumptions of small overpotentials and weak side reaction, which are reasonable in the vast majority of practical scenarios, including SEI growth, lithium plating and both mechanisms combined.


We found that the SPMe+SR retains most of the accuracy of the DFN+SR but it is one order of magnitude simpler and significantly faster (over 150 times faster in some cases, but strongly dependent on solver and mesh size). The simulations of the SPMe+SR have been compared against those from DFN+SR, for three scenarios (SEI growth, lithium plating, and both effects combined) showing that the former can capture accurately the various features of degradation (capacity fade, voltage profile and porosity distribution) at the two time scales of the system (a single cycle and battery lifetime). The main limitation of the SPMe+SR is that it cannot capture electrolyte depletion because in that situation some particles are lithiated and some are not, breaking down the homogeneous electrode behaviour that the SPMe+SR describes. Apart from presenting the reduced model, this article also provides a formal \rv{derivation framework} using asymptotic methods, which can be applied to models including different degradation mechanisms. The SPMe+SR provides similar results to the DFN+SR, which is the state-of-the-art in terms of physics-based models for battery degradation, but with a much lower complexity. This means that the model is simpler (i.e. fewer and simpler equations) which makes it easier to analyse, understand and implement, and faster to simulate. All these advantages make the SPMe+SR suitable for demanding battery applications, such as design, control and diagnosis. In particular, given that it has the structure of a Single Particle Model, existing techniques for such models can be used (e.g. develop an observer for control use). Future areas of research include validating the SPMe+SR against experimental data, extending it to account for electrolyte depletion, and developing techniques based on the reduced model for parameterisation of degradation models.

\section*{Data and code availability}
The code to reproduce the results presented in this article is publicly available on the repository: \\
\url{https://www.github.com/brosaplanella/SPMe_SR}
(DOI: 10.5281/zenodo.6624983).

A checklist detailing key aspects of the model \cite{Mistry2021} is provided in the Supplementary Information. 

\section*{Acknowledgements}
This work is supported by The Faraday Institution ``Multi-Scale Modelling'' project [EP/S003053/1 grant numbers FIRG003 and FIRG025] and the Innovate UK  ``COBRA: Cloud/On-board Battery Remaining useful life Algorithm'' project [TSB number 100831]. The authors would like to thank the team at Eatron Technologies for the useful discussions.

\bibliographystyle{elsarticle-num}
\bibliography{references}

\appendix

\section{Doyle-Fuller-Newman model with Side Reactions (DFN+SR)}\label{sec:DFN+SR}
In this section we state the dimensional model of the Doyle-Fuller-Newman model with Side Reactions (DFN+SR) and derive its dimensionless form. This dimensionless form is the starting point of the asymptotic reduction presented in Section \ref{sec:derivation_SPMe_SR}. We include the side reaction on the negative electrode only, but note that the model could be easily extended to include side reactions in the positive electrode as well. The parameters and variables involved in the model are introduced in Section~\ref{sec:SPMe+SR}.

\rv{This specific formulation of DFN+SR is based on the model introduced in \cite{Yang2017}. It has been widely used in the literature and shown good agreement with experimental data, and that is why it was chosen as the starting point of this analysis. In particular, the porosity equation is built on the assumption of a thin film so the the film growth is related to the porosity change as $\pdv{\varepsilon_k}{t} = - a_k \pdv{L_{\mathrm{f}, k}}{t}$ which corresponds to the planar case. This assumption might not be reasonable at the later stages of aging and a more realistic model should be derived, but this is out of the scope of this article. However, the analysis presented in this article would still be valid for other porosity models.}

\subsection{Dimensional model}\label{sec:DFN+SR_dim}
In each particle, the lithium concentration is governed by
\begin{subequations}
\begin{align}
\pdv{c_{k}}{t} &= \frac{1}{r^2} \pdv{}{r} \left(r^2 D_{k} \pdv{c_{k}}{r} \right), & \quad \text{ in } 0 < r < R_k,\\
\pdv{c_{k}}{r} &= 0, & \quad \text{ at } r = 0,\\
- D_{k} \pdv{c_{k}}{r} &= \frac{J_{k,\mathrm{int}}}{a_k F}, & \quad \text{ at } r = R_k,\\
c_{k} &= c_{k, \mathrm{init}}, & \quad \text{ at } t = 0,
\end{align}
and the potential in each electrode is described by
\begin{align}
\pdv{i_{k}}{x} &= -J_k,\\
i_{k} &= -\sigma_{k} \pdv{\phi_{k}}{x},
\end{align}
In both cases, the $x$ domain is defined to be $0 \leq x \leq L_\mrn$ if $k = \mrn$, and $L - L_\mrp \leq x \leq L$ if $k = \mrp$.

The electrolyte equations, which are defined in the domain $0 \leq x \leq L$, are
\begin{align}
\pdv{}{t} \left( \varepsilon c_\mre \right) &= - \pdv{N_{\mre}}{x} + \frac{J}{F},\\
\pdv{i_{\mre}}{x} &= J,
\end{align}
with
\begin{align}
N_{\mre} &= -D_\mre \mathcal{B}(x) \pdv{c_{\mre}}{x} + t^+ \frac{i_{\mre}}{F},\\
i_{\mre} &= -\sigma_{\mre} \mathcal{B}(x) \left(\pdv{\phi_{\mre}}{x} - 2(1-t^+) \left(1 + \pdv{f_\pm}{c_\mre} \right) \frac{R T}{F} \pdv{\log c_{\mre}}{x} \right).
\end{align}
\end{subequations}

The intercalation reaction between the electrode and the electrolyte is given by
\begin{subequations}
\begin{align}
J(x) &= \begin{cases}
J_\mrn = J_{\mrn,\mathrm{int}} + J_\SR, & \text{ for } 0 \leq x \leq L_\mrn,\\
0, & \text{ for } L_\mrn < x \leq L - L_\mrp,\\
J_\mrp = J_{\mrp,\mathrm{int}}, & \text{ for } L - L_\mrp < x \leq L,
\end{cases},
\end{align}
where the intercalation reaction kinetics are driven by
\begin{align}
J_{k,\mathrm{int}} &= a_k m_k \left. \sqrt{c_{\mre} c_{k} \left(c_{k}^{\max} - c_{k} \right)} \right|_{r = R_k} \sinh \left(\frac{1}{2} \frac{F}{R T} \eta_k \right),\\
\eta_k &= \phi_{k} - \phi_{\mre} - U_k\left( \left. c_{k} \right|_{r = R_k} \right) - J_k \frac{L_{\mathrm{f}, k}}{a_k \sigma_{\mathrm{f}, k}},
\end{align}
and the side reaction kinetics are driven by
\begin{align}
J_\SR &= - a_\mrn j_\SR \exp \left(- \alpha_\SR \frac{F}{R T} \eta_\SR \right),\\
\eta_\SR &= \phi_{\mrn} - \phi_{\mre} - U_\SR - J_\mrn \frac{L_{\mathrm{f},\mrn}}{a_\mrn \sigma_{\mathrm{f},\mrn}},
\end{align}
\end{subequations}
where $j_\SR$ is the side reaction exchange current density which might involving solving additional equations (e.g. the SEI models in \cite{Atalay2020,Yang2017}).

This side reaction causes a change in the porosity of the negative electrode that follows
\begin{equation}
    \pdv{\varepsilon}{t} = \frac{M_\SR}{n_\SR \rho_\SR F} J_\SR,
\end{equation}
and the porosity across the cell is defined as
\begin{equation}
    \varepsilon (x, t) = 
    \begin{cases}
    \varepsilon_\mrn (x, t), & \text{ if } 0 \leq x < L_\mrn,\\
    \varepsilon_\mrs (x), & \text{ if } L_\mrn \leq x < L - L_\mrp,\\
    \varepsilon_\mrp (x), & \text{ if } L - L_\mrp \leq x \leq L.
    \end{cases}
\end{equation}

Note that the porosity can be related to the film thickness as
\begin{equation}
    L_{\mathrm{f},k} = L_{\mathrm{f},k, \mathrm{init}} - \frac{1}{a_k} (\varepsilon_k - \varepsilon_{k, \mathrm{init}}),
\end{equation}
so we could write an ODE for film thickness instead of this one for the porosity.

The boundary conditions are the following. At the current collector ends we impose
\begin{subequations}
\begin{align}
i_{\mrn} &= i_\mathrm{app}, & N_{\mre} &= 0, & \phi_{\mre} &= 0, & \text{ at } x &= 0,\\
i_{\mrp} &= i_\mathrm{app}, & N_{\mre} &= 0, & i_{\mre} &= 0, & \text{ at } x &= L.
\end{align}

At the electrode-separator interfaces, we impose zero current in the electrodes
\begin{align}
i_{\mrn} &= 0, & \text{ at } x = L_\mrn,\\
i_{\mrp} &= 0, & \text{ at } x = L - L_\mrp.
\end{align}

Finally, we impose the initial condition for the electrolyte concentration
\begin{align}
c_{\mre} &= c_{\mre,\mathrm{init}}, & \text{ at } t = 0.
\end{align}
\end{subequations}

\subsection{Dimensionless model}\label{sec:DFN+SR_ND}
We define the following scalings of the problem
\begin{equation}\label{eq:scalings}
\begin{aligned}
t &= t_0 \hat{t}, & c_{k} &= c_{k}^{\max} \hat{c}_{k}, & \phi_{k} &= \phi_0 \hat{\phi}_{k}, & i_{k} &= i_0 \hat{i}_{k}, & J_k &= \frac{i_0}{L} \hat{J}_k, \\ 
x &= L \hat{x}, & c_\mre &= c_{\mre,\mathrm{init}} \hat{c}_\mre, & \phi_{\mre} &= \frac{R T}{F} \hat{\phi}_{\mre}, & i_{\mre} &= i_0 \hat{i}_{\mre} & j_k &= \frac{i_0}{a_k L} \hat{j}_k, \\
r_k &= R_k \hat{r}_k, & D_k &= D_{k,\typ} \hat{D}_k, & \eta_k &= \frac{R T}{F} \hat{\eta}_k, & i_\mathrm{app} &= i_0 \hat{i}_\mathrm{app}, & L_{\mathrm{f}, k} &= L_{\mathrm{f}, k, \mathrm{init}} \hat L_{\mathrm{f}, k}, \\
\sigma_\mre &= \sigma_{\mre,\typ} \hat{\sigma}_\mre, & D_\mre &= D_{\mre,\typ} \hat{D}_\mre, & U_k &= \phi_0 \hat{U}_k, & N_{\mre} &= \frac{D_{\mre,\typ} c_{\mre,\mathrm{init}}}{L} \hat{N}_{\mre}, & \\
\end{aligned}
\end{equation}
and we choose the time scale $t_0$ to be the discharge time scale
\begin{equation}
    t_0 = \frac{F c_{\mrn}^{\max} L}{i_0}.
\end{equation}

The parameters $i_0$ and $\phi_0$ are the typical current and electrode potential, respectively, and the subscript typ denotes the typical value of that parameter. \rv{Note that if the initial film thickness is zero, we could take a typical film thickness $L_{\mathrm{f}, k, \mathrm{typ}}$ as a scaling and the analysis would still remain valid.}

Then, we can write the dimensionless model as follows. The model for concentration in the particles reads
\begin{subequations}\label{eq:DFN+SR_ND}
\begin{align}
\mathcal{C}_k \pdv{\hat c_{k}}{\hat t} &= \frac{1}{\hat r^2} \pdv{}{\hat r} \left(\hat r^2 \hat D_{k} \pdv{\hat c_{k}}{\hat r} \right), & \quad \text{ in } 0 < \hat r < 1,\\
\pdv{\hat c_{k}}{\hat r} &= 0, & \quad \text{ at } \hat r = 0,\\
- \hat D_{k} \pdv{\hat c_{k}}{\hat r} &= \frac{\mathcal{C}_k}{\alpha_k \gamma_k} \hat J_{k,\mathrm{int}}, & \quad \text{ at } \hat r = 1,\\
\hat c_{k} &= \mu_k, & \quad \text{ at } \hat t = 0,
\end{align}
and the potential in each electrode is given by
\begin{align}
\pdv{\hat i_{k}}{\hat x} &= -\hat J_k,\\
\hat i_{k} &= - \lambda \Sigma_{k} \pdv{\hat \phi_{k}}{\hat x},
\end{align}
Now, the $\hat x$ domain is defined to be $0 \leq \hat x \leq \ell_\mrn$ if $k = \mrn$, and $1 - \ell_\mrp \leq \hat x \leq 1$ if $k = \mrp$.

The electrolyte equations, which are defined in the domain $0 \leq \hat x \leq 1$, are
\begin{align}
\mathcal{C}_\mre \gamma_\mre \pdv{}{\hat t} \left( \varepsilon \hat c_\mre \right) &= - \gamma_\mre \pdv{\hat N_{\mre}}{\hat x} + \mathcal{C}_\mre \hat J,\\
\pdv{\hat i_{\mre}}{\hat x} &= \hat J,
\end{align}
with
\begin{align}
\hat N_{\mre} &= -\hat D_\mre \mathcal{B}(\hat x) \pdv{\hat c_{\mre}}{\hat x} + t^+ \frac{\mathcal{C}_\mre}{\gamma_\mre} \hat i_\mre,\\
\hat i_{\mre} &= -\Sigma_\mre \hat \sigma_{\mre} \mathcal{B}(\hat x) \left(\pdv{\hat \phi_{\mre}}{\hat x} - 2(1-t^+) \left(1 + \pdv{f_\pm}{c_\mre} \right) \pdv{\log \hat c_{\mre}}{\hat x} \right).
\end{align}
\end{subequations}

The intercalation reaction between the electrode and the electrolyte is given by
\begin{subequations}\label{eq:DFN+SR_ND_reaction}
\begin{align}
\hat J &= \begin{cases}
\hat J_\mrn = \hat J_{\mrn,\mathrm{int}} + \hat J_\SR, & \text{ if } 0 \leq \hat x \leq \ell_\mrn,\\
0, & \text{ if } \ell_\mrn < \hat x \leq 1 - \ell_\mrp,\\
\hat J_\mrp = \hat J_{\mrp,\mathrm{int}}, & \text{ if } 1 - \ell_\mrp < \hat x \leq 1,
\end{cases}
\end{align}
where the intercalation reaction kinetics are driven by
\begin{align}
\hat J_{k,\mathrm{int}} &= \frac{\gamma_k}{\mathcal{C}_{\mathrm{r},k}} \left. \sqrt{\hat c_{\mre} \hat c_{k} \left(1 - \hat c_{k} \right)} \right|_{\hat r = 1} \sinh \left(\frac{\hat \eta_k}{2}\right),\\
\hat \eta_k &= \lambda \left( \hat \phi_{k} - \hat U_k\left( \left. \hat c_{k} \right|_{\hat r = 1} \right) \right) - \hat \phi_{\mre} - \frac{\hat J_k \hat L_{\mathrm{f}, k}}{\Sigma_{\mathrm{f}, k}},
\end{align}
and the side reaction kinetics are driven by
\begin{align}
\hat J_\SR &= - \hat j_\SR \exp \left(- \alpha_\SR \hat \eta_\SR \right),\\
\hat \eta_\SR &= \lambda \left( \hat \phi_{\mrn} - \hat U_\SR \right) - \hat \phi_{\mre}  - \frac{\hat J_\mrn \hat L_{\mathrm{f},\mrn}}{\Sigma_{\mathrm{f},\mrn}}.
\end{align}
\end{subequations}
This side reaction causes a change in the porosity of the negative electrode that follows
\begin{equation}\label{eq:DFN+SR_ND_porosity}
    \pdv{\varepsilon_\mrn}{\hat t} = \frac{1}{n_\SR \gamma_\SR} \hat J_\SR,
\end{equation}
and the corresponding film thickness can be calculated as
\begin{equation}
    \hat L_{\mathrm{f},k} = 1 - \frac{1}{\beta_k} (\varepsilon_k - \varepsilon_{k, \mathrm{init}}).
\end{equation}

The boundary conditions at current collector ends are
\begin{subequations}\label{eq:DFN+SR_ND_BC}
\begin{align}
\hat i_{\mrn} &= \hat i_\mathrm{app}, & \hat N_{\mre} &= 0, & \hat \phi_{\mre} &= 0, & \text{ at } \hat x &= 0,\\
\hat i_{\mrp} &= \hat i_\mathrm{app}, & \hat N_{\mre} &= 0, & \hat i_{\mre} &= 0, & \text{ at } \hat x &= 1,
\end{align}
and at the electrode-separator interfaces are
\begin{align}
\hat i_{\mrn} &= 0, & \text{ at } \hat x = \ell_\mrn,\\
\hat i_{\mrp} &= 0, & \text{ at } \hat x = 1 - \ell_\mrp.
\end{align}
Finally, the initial condition for the electrolyte is
\begin{align}
\hat c_{\mre} &= 1, & \text{ at } \hat t = 0.
\end{align}
\end{subequations}

The dimensionless parameters of the model are
\begin{equation}\label{eq:ND_parameters}
\begin{aligned}
\mathcal{C}_k &= \frac{R_k^2}{D_{k,\typ} t_0}, & \mathcal{C}_\mre &= \frac{L^2}{D_{\mre,\mathrm{typ}} t_0}, & \mathcal{C}_{\mathrm{r}, k} &= \frac{F}{m_k a_k \sqrt{c_{\mre,\mathrm{init}}} t_0}, & \mu_k &= \frac{c_{k, \mathrm{init}}}{c_{k}^{\max}}, & \alpha_k &= a_k R_k,\\ 
\gamma_k &= \frac{c_{k}^{\max}}{c_{\mrn}^{\max}}, & \gamma_\mre &= \frac{c_{\mre,\mathrm{init}}}{c_{\mrn}^{\max}}, & \gamma_\SR &= \frac{\rho_\SR}{M_\SR c_\mrn^{\max}}, & \lambda &= \frac{\phi_0 F}{R T}, & \beta_k &= a_k L_{\mathrm{f},k, \mathrm{init}} \\
\Sigma_{k} &= \frac{R T}{F L i_0} \sigma_{\mrs,k}, & \Sigma_{\mre} &= \frac{R T}{F L i_0} \sigma_{\mre, \mathrm{typ}}, & \Sigma_{\mathrm{f}, k} &= \frac{R T}{F L_{\mathrm{f}, k, \mathrm{init}} i_0} \sigma_{\mathrm{f}, k} a_k L, & \ell_k &= \frac{L_k}{L}.
\end{aligned}
\end{equation}

The typical values of these parameters are shown in Table \ref{tab:parameter_values_ND}.

\begin{table}
\centering
\begin{tabular}{| c c c |}
\hline
\textbf{Symbol} & \textbf{Pos.} & \textbf{Neg.} \\ \hline
$\mathcal{C}_k$ & $0.60C$ & $9.17\E{-2}C$ \\
$\mathcal{C}_{\mathrm{r},k}$ & $0.21C$ & $1.08C$ \\
$\mu_k$ & $0.27$ & $0.90$ \\
$\alpha_k$ & $2.00$ & $2.25$ \\
$\gamma_k$ & $1.90$ & $1$ \\
$\Sigma_k$ & $0.55C^{-1}$ & $656.18C^{-1}$ \\
$\Sigma_{\mathrm{f},k}$ & - & $35C^{-1}$ \\
$\beta_k$ & - & $2.0\E{-3}$ \\
$\ell_k$ & $0.44$ & $0.49$ \\ \hline
$\mathcal{C}_\mre$ & \multicolumn{2}{c |}{$1.49\E{-2}C$} \\
$\gamma_\mre$ & \multicolumn{2}{c |}{$3.02\E{-2}$} \\
$\Sigma_\mre$ & \multicolumn{2}{c |}{$2.90C^{-1}$} \\ \hline
$\gamma_\SR$ & \multicolumn{2}{c |}{$0.31$ / $2.32$} \\
$\lambda$ & \multicolumn{2}{c |}{$38.94$} \\\hline
\end{tabular}
\caption{Dimensionless parameters for the model, calculated from the dimensional parameters in Tables \ref{tab:parameter_values_LGM50} and \ref{tab:parameter_values_SR}. Even though the model allows for an SEI layer on the positive electrode, in the simulations we do not account for them so we do not have parameter values for $\Sigma_{\mathrm{f},\mrp}$ and $\beta_\mrp$. The two values for $\gamma_\SR$ correspond to SEI and lithium plating, respectively.}
\label{tab:parameter_values_ND}
\end{table}

\section{Dimensionless form of SR models}\label{sec:SR_models_ND}
Even though it is not necessary for the analysis, for completion we present here the dimensionless form of the examples presented in Section \ref{sec:SR_models}. Using the scaling in \eqref{eq:scalings} and defining $\hat c_\SEI = c_{\SEI, \mathrm{init}} \hat c_\SEI$, we find that the SEI growth model is
\begin{equation}\label{eq:j_SEI_ND}
    \hat j_\SEI = \frac{1}{\mathcal{C}_{\mathrm{r},\SEI}} \hat c_\SEI,
\end{equation}
with
\begin{equation}\label{eq:c_SEI_ND}
    \hat c_\SEI = 1 + \mathcal{C}_\SEI \hat j_\SEI \hat L_\mathrm{f},
\end{equation}
and
\begin{equation}\label{eq:j_Li_ND}
    \hat j_\Li = \frac{1}{\mathcal{C}_{\mathrm{r},\Li}} \hat c_\mre,
\end{equation}
where the dimensionless parameters have been defined as 
\begin{align}
    \mathcal{C}_{\mathrm{r},\SEI} &= \frac{c_\mrn^{\max}}{a_\mrn k_\SEI t_0 c_{\SEI,\mathrm{init}}} \approx 1675 C, & \mathcal{C}_\SEI &= \frac{L_{\mathrm{f},\mathrm{init}} c_\mrn^{\max}}{t_0 a_\mrn D_\SEI c_{\SEI,\mathrm{init}}} \approx 41.9 C, & \mathcal{C}_{\mathrm{r},\Li} &= \frac{c_\mrn^{\max}}{a_\mrn k_\Li t_0 c_{\mre,\mathrm{init}}} \approx 761 C.
\end{align}
The values have been calculated from the parameter values in Tables \ref{tab:parameter_values_LGM50} and \ref{tab:parameter_values_SR}.

\rv{\section{Dimensionless form of SPMe+SR}\label{sec:model_summary}}
In this section we present the dimensionless form of the SPMe+SR for completeness, by combining the results of the derivation presented in Section \ref{sec:asymptotic_reduction}. Note that, for the particle diffusivity and the open-circuit potentials, we assumed the functions to be their first order asymptotic expansion, that is
\begin{equation}
\begin{aligned}
    D_k (\bar c_k) &\approx D_k (\bar c_{k0}) + \lambda^{-1} D_k' (\bar c_{k0}) \bar c_{k1},\\
    U_k (\bar c_k) &\approx U_k (\bar c_{k0}) + \lambda^{-1} U_k' (\bar c_{k0}) \bar c_{k1}.
\end{aligned}
\end{equation}

\paragraph{Particle equations} The equation for the concentration in the negative averaged particle reads
\begin{subequations}
\begin{align}
\mathcal{C}_\mrn \pdv{\bar c_\mrn}{t} &= \frac{1}{r^2} \pdv{}{r} \left(r^2 D_\mrn(\bar c_\mrn) \pdv{\bar c_\mrn}{r} \right), & \quad \text{ in } 0 < r < 1,\\
\pdv{\bar c_\mrn}{r} &= 0, & \quad \text{ at } r = 0,\\
- D_\mrn (\bar c_\mrn) \pdv{\bar c_\mrn}{r} &= \frac{\mathcal{C}_\mrn}{\alpha_\mrn \gamma_\mrn} \left( \frac{i_\mathrm{app}}{\ell_\mrn} - \bar J_\SR \right), & \quad \text{ at } r = 1,\\
\bar c_\mrn &= \mu_\mrn, & \quad \text{ at } t = 0,
\end{align}
\end{subequations}
where
\begin{subequations}
\begin{align}
    \bar J_\SR &= \frac{1}{\ell_\mrn} \int_0^{\ell_\mrn} J_\SR \dd x,\\
    J_\SR &= - j_{\SR} \exp \left( - \alpha_\SR \left( \lambda (\phi_{\mrn} - U_\SR) - \phi_{\mre } - \frac{i_\mathrm{app} L_{\mathrm{f},\mrn}}{\ell_\mrn \Sigma_{\mathrm{f},\mrn}}\right) \right).
\end{align}
\end{subequations}

The positive electrode averaged particle equation reads
\begin{subequations}
\begin{align}
\mathcal{C}_\mrp \pdv{\bar c_\mrp}{t} &= \frac{1}{r^2} \pdv{}{r} \left(r^2 D_\mrp( \bar c_\mrp) \pdv{\bar c_\mrp}{r} \right), & \quad \text{ in } 0 < r < 1,\\
\pdv{\bar c_\mrp}{r} &= 0, & \quad \text{ at } r = 0,\\
- D_\mrp (\bar c_\mrp) \pdv{\bar c_\mrp}{r} &= - \frac{\mathcal{C}_\mrp}{\alpha_\mrp \gamma_\mrp} \frac{i_\mathrm{app}}{\ell_\mrp}, & \quad \text{ at } r = 1,\\
\bar c_\mrp &= \mu_\mrp. & \quad \text{ at } t = 0,
\end{align}
\end{subequations}

\paragraph{Electrolyte equation} The equation for the electrolyte concentration reads
\begin{align}
    \mathcal{C}_\mre \gamma_\mre \pdv{}{t} \left( \varepsilon c_{\mre} \right) &=  \pdv{}{x} \left(\gamma_\mre D_\mre (c_{\mre}) \mathcal{B}_(x) \pdv{c_{\mre}}{x} + (1 - t^+(c_{\mre})) \mathcal{C}_\mre i_{\mre} \right), & \text{ in } 0 \leq x \leq 1,\\
    \pdv{c_{\mre}}{x} &= 0, & \text{ at } x = 0,1,\\
    c_{\mre} &= 1, & \text{ at } t = 0,
\end{align}
with
\begin{align}
    i_{\mre} &= \begin{cases}
    \frac{i_\mathrm{app}}{\ell_\mrn}x, & \text{ if } 0 \leq x < \ell_\mrn,\\
    i_\mathrm{app}, & \text{ if } \ell_\mrn \leq x < 1 - \ell_\mrp,\\
    \frac{i_\mathrm{app}}{\ell_\mrp} (1 - x), & \text{ if } 1 - \ell_\mrp \leq x \leq 1.
    \end{cases}
\end{align}

\paragraph{Porosity variation} The equation for the variation of the porosity is
\begin{equation}
    \pdv{\varepsilon_\mrn}{t} = \frac{J_\SR}{n_\SR \gamma_\SR},
\end{equation}
and the film thickness can be calculated as
\begin{equation}
    L_{\mathrm{f}, k} = 1 - \frac{1}{\beta_k} \left( \varepsilon_k - \varepsilon_{k, \mathrm{init}} \right).
\end{equation}

\paragraph{Expressions for the potentials} The potentials in the electrodes and electrolyte read
\begin{equation}
\begin{aligned}
    \phi_\mrn &= U_\mrn \left( \left. \bar c_\mrn \right|_{r = 1} \right) + \lambda^{-1} \left( - \frac{i_\mathrm{app} (2 \ell_\mrn - x) x}{2 \ell_\mrn \Sigma_\mrn} + \frac{i_\mathrm{app} \ell_\mrn}{3 \Sigma_\mrn} + \frac{i_\mathrm{app} \bar L_{\mathrm{f}, \mrn}}{\ell_\mrn \Sigma_{\mathrm{f},\mrn}} - \frac{1}{\ell_\mrn \Sigma_\mre} \int_0^{\ell_\mrn} \int_0^x \frac{i_{\mre}(s,t) \dd s}{\sigma_\mre (c_{\mre}(s,t)) \mathcal{B}(s, t)} \dd x \right. \\
    & \left. \quad + \frac{1}{\ell_\mrn} \int_0^{\ell_\mrn} \int_0^x 2 (1 - t^+(c_{\mre}(s, t))) \left(1 + \pdv{f_\pm}{c_{\mre}} \right) \pdv{\log c_{\mre}(s,t)}{s} \dd s \dd x + \frac{2}{\ell_\mrn} \int_0^{\ell_\mrn} \arcsinh\left(\frac{i_\mathrm{app}}{\ell_\mrn j_{\mrn}} \right) \dd x \right),\\
    \phi_\mrp &= U_\mrp \left( \left. \bar c_\mrp \right|_{r = 1} \right) + \lambda^{-1} \left( \frac{i_\mathrm{app} (2 (1 - \ell_\mrp) - x) x}{2 \ell_\mrp \Sigma_\mrp} - \frac{i_\mathrm{app}(2 \ell_\mrp^2 - 6 \ell_\mrp + 3)}{6 \ell_\mrp \Sigma_\mrp} - \frac{i_\mathrm{app} \bar L_{\mathrm{f}, \mrp}}{\ell_\mrp \Sigma_{\mathrm{f},\mrp}} \right. \\ 
    & \quad - \frac{1}{\ell_\mrp \Sigma_\mre} \int_{1 - \ell_\mrp}^1 \int_0^x \frac{i_{\mre}(s,t) \dd s}{\sigma_\mre (c_{\mre}(s,t)) \mathcal{B}(s)} \dd x \\
    & \left. \quad + \frac{1}{\ell_\mrp} \int_{1 - \ell_\mrp}^1 \int_0^x 2 (1 - t^+(c_{\mre}(s, t))) \left(1 + \pdv{f_\pm}{c_{\mre}} \right) \pdv{\log c_{\mre}(s,t)}{s} \dd s \dd x - \frac{2}{\ell_\mrp} \int_{1 - \ell_\mrp}^1 \arcsinh\left(\frac{i_\mathrm{app}}{\ell_\mrp j_{\mrp}} \right) \dd x \right),\\
    \phi_\mre &= - \int_0^x \frac{i_{\mre}}{\Sigma_\mre \sigma_\mre (c_{\mre}(s, t)) \mathcal{B}(s)} \dd s + \int_0^x 2 (1 - t^+(c_{\mre}(s, t))) \left(1 + \pdv{f_\pm}{c_{\mre}} \right) \pdv{\log c_{\mre}(s,t)}{s} \dd s.
\end{aligned}
\end{equation}
where
\begin{align}
    j_k &= \frac{\gamma_k}{\mathcal{C}_{\mathrm{r},k}} \left. \sqrt{c_{\mre} \bar c_{k} \left(1 - \bar c_{k} \right)} \right|_{r = 1}.
\end{align}
Note that, to be precise, in the exchange current density we should have $\bar c_{k 0}$ instead of $\bar c_k$. However, the contribution of $\bar c_{k 1}$ is very small so this is a reasonable assumption. This correction would be included in the next correction to the voltage, but that would require solving additional equations for the potentials and the electrolyte concentration.

Finally, the terminal voltage can be calculated by taking
\begin{equation}
    V = \left. \phi_\mrp \right|_{x = 1} - \left. \phi_\mrn \right|_{x = 0}.
\end{equation}
We can write this expression as
\begin{equation}
    V = U_\mathrm{eq} + \eta_\mathrm{r} + \eta_\mre + \Delta \phi_\mre + \Delta \phi_\mrs + \Delta \phi_\mathrm{f},
\end{equation}
where
\begin{equation}
\begin{aligned}
U_\mathrm{eq} &= U_\mrp \left( \left. \bar c_\mrp \right|_{r = 1} \right) - U_\mrn \left( \left. \bar c_\mrn \right|_{r = 1} \right), \\
\eta_\mathrm{r} &= - 2 \lambda^{-1} \left(\frac{1}{\ell_\mrp} \int_{1 - \ell_\mrp}^1 \arcsinh\left(\frac{i_\mathrm{app}}{\ell_\mrp j_{\mrp}} \right) \dd x + \frac{1}{\ell_\mrn} \int_0^{\ell_\mrn} \arcsinh\left(\frac{i_\mathrm{app}}{\ell_\mrn j_{\mrn}} \right) \dd x \right), \\
\eta_\mre &= \lambda^{-1} \left(\frac{1}{\ell_\mrp} \int_{1 - \ell_\mrp}^1 \int_0^x 2 (1 - t^+(c_{\mre}(s, t))) \left(1 + \pdv{f_\pm}{c_{\mre}} \right) \pdv{\log c_{\mre}(s,t)}{s} \dd s \dd x \right. \\
& \left. \quad - \frac{1}{\ell_\mrn} \int_0^{\ell_\mrn} \int_0^x 2 (1 - t^+(c_{\mre}(s, t))) \left(1 + \pdv{f_\pm}{c_{\mre}} \right) \pdv{\log c_{\mre}(s,t)}{s} \dd s \dd x \right), \\
\Delta \phi_\mre &= \lambda^{-1} \left( - \frac{1}{\ell_\mrp \Sigma_\mre} \int_{1 - \ell_\mrp}^1 \int_0^x \frac{i_{\mre}(s,t) \dd s}{\sigma_\mre (c_{\mre}(s,t)) \mathcal{B}(s)} \dd x + \frac{1}{\ell_\mrn \Sigma_\mre} \int_0^{\ell_\mrn} \int_0^x \frac{i_{\mre}(s,t) \dd s}{\sigma_\mre (c_{\mre}(s,t)) \mathcal{B}(s, t)} \dd x  \right), \\
\Delta \phi_\mrs &= - \lambda^{-1} \frac{i_\mathrm{app}}{3} \left( \frac{\ell_\mrp}{\Sigma_\mrp} + \frac{\ell_\mrn}{\Sigma_\mrn} \right), \\
\Delta \phi_\mathrm{f} &= - \lambda^{-1} i_\mathrm{app} \left( \frac{\bar L_{\mathrm{f},\mrp}}{\ell_\mrp \Sigma_{\mathrm{f}, \mrp}} + \frac{\bar L_{\mathrm{f},\mrn}}{\ell_\mrn \Sigma_{\mathrm{f}, \mrn}} \right).
\end{aligned}
\end{equation}

The redimensionalised form of this model is the one provided in Section \ref{sec:SPMe+SR}.

\section{Parameter values}
In this section we provide details on the parameters that are functions, which were too long to be included in Table \ref{tab:parameter_values_LGM50}. These parameters are taken from \cite{Chen2020} (but note that the electrolyte parameters come from \cite{Nyman2008}).

\rv{The ion diffusivity in the electrolyte, measured in $\mathrm{m}^2 \; \mathrm{s}^{-1}$, is given by
\begin{equation}\label{eq:D_e}
    D_\mre(c_\mre) = 8.794 \cdot 10^{-17} c_\mre ^ 2 - 3.972 \cdot 10 ^ {-13} c_\mre + 4.862 \cdot 10 ^ {-10}.
\end{equation}
while the conductivity of the electrolyte, measured in $\mathrm{S} \; \mathrm{m}^{-1}$, is given by
\begin{equation}\label{eq:sigma_e}
    \sigma_\mre (c_\mre) = 1.297 \cdot 10^{-10} c_\mre ^ 3 - 7.937 \cdot 10^{-5} c_\mre ^{1.5} + 3.329 \cdot 10^{-3} c_\mre.
\end{equation}
Note the change in the values compared to \cite{Nyman2008}, as here we have defined the ion concentration in the electrolyte in $\mathrm{mol} \; \mathrm{m}^{-3}$.}

The open-circuit potentials for the positive and negative electrodes, respectively, as a function of the stoichiometry are
\begin{subequations}
\begin{multline}\label{eq:U_p}
    U_\mrp(x) = -0.8090 x + 4.4875 - 0.0428 \tanh{\left(18.5138 (x - 0.5542) \right)} \\ - 17.7326 \tanh{\left( 15.7890 (x - 0.3117) \right)} + 17.5842 \tanh{\left( 15.9308 (x - 0.3120) \right)},
\end{multline}
\begin{multline}\label{eq:U_n}
    U_\mrn(x) = 1.9793 \exp \left( -39.3631 x \right) + 0.2482 - 0.0909 \tanh \left(29.8538 (x - 0.1234) \right) \\ - 0.04478 \tanh \left( 14.9159 (x - 0.2769) \right) - 0.0205 \tanh \left( 30.4444 (x - 0.6103) \right).
\end{multline}
\end{subequations}

\end{document}


\maketitle

\tableofcontents

\section{Additional simulations}
To validate the reduced model we performed the simulations for the four possible combinations of charge rates (C/3 and C/2) and discharge rates (1C and 2C). In the main body of the article we only showed C/2 charge -- 1C discharge as it was representative enough, but for completeness here we provide the other simulations. The code to reproduce these results is available at \url{https://www.github.com/brosaplanella/SPMe_SR}.

\subsection{SEI growth}
First we show the results for SEI growth only.

\begin{figure}[!htb]
    \centering
    \includegraphics[scale=1]{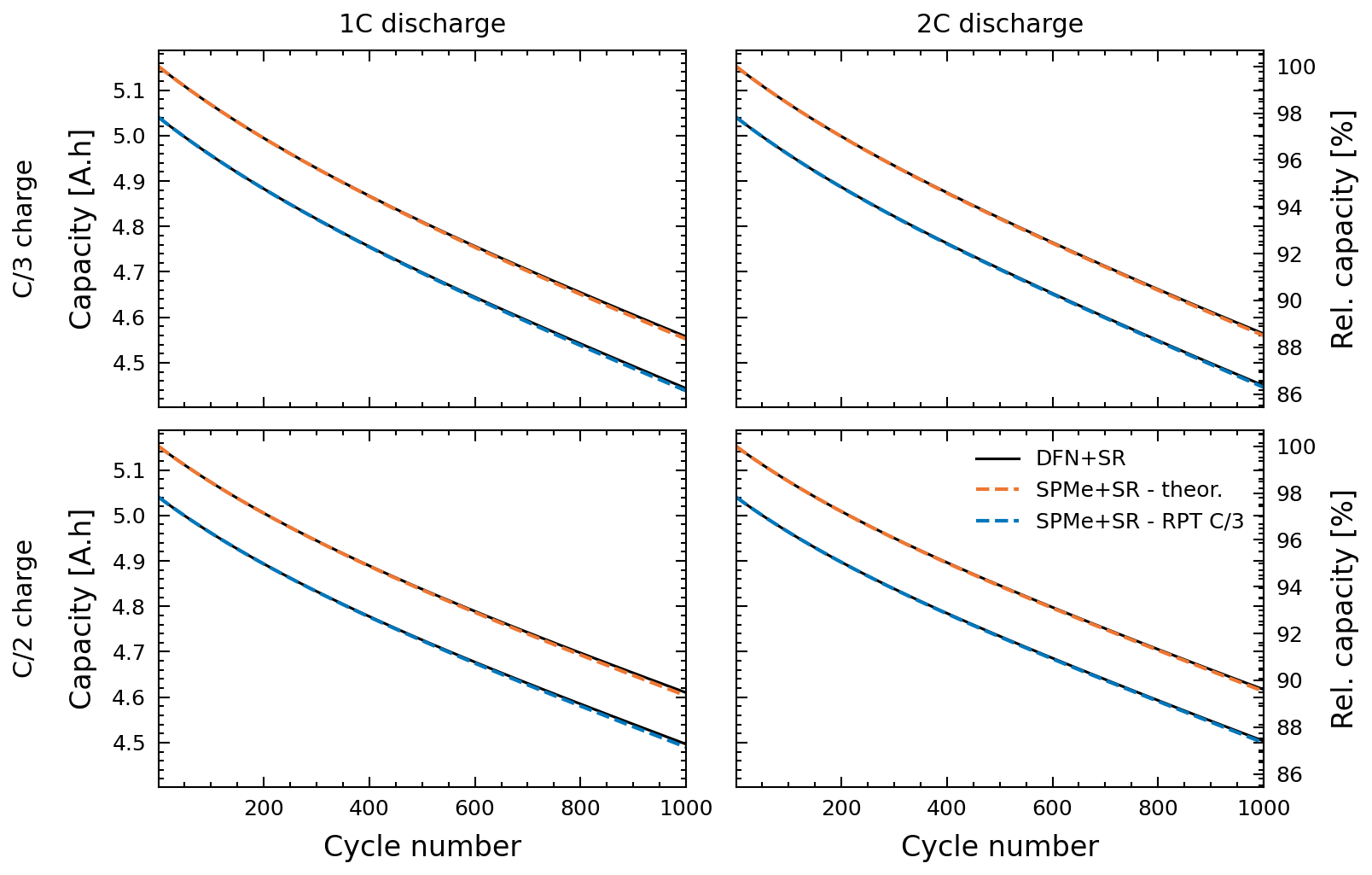}
    \caption{Compare capacity for the model with SEI growth and porosity change.}
\end{figure}

\begin{figure}[!htb]
    \centering
    \includegraphics[scale=1]{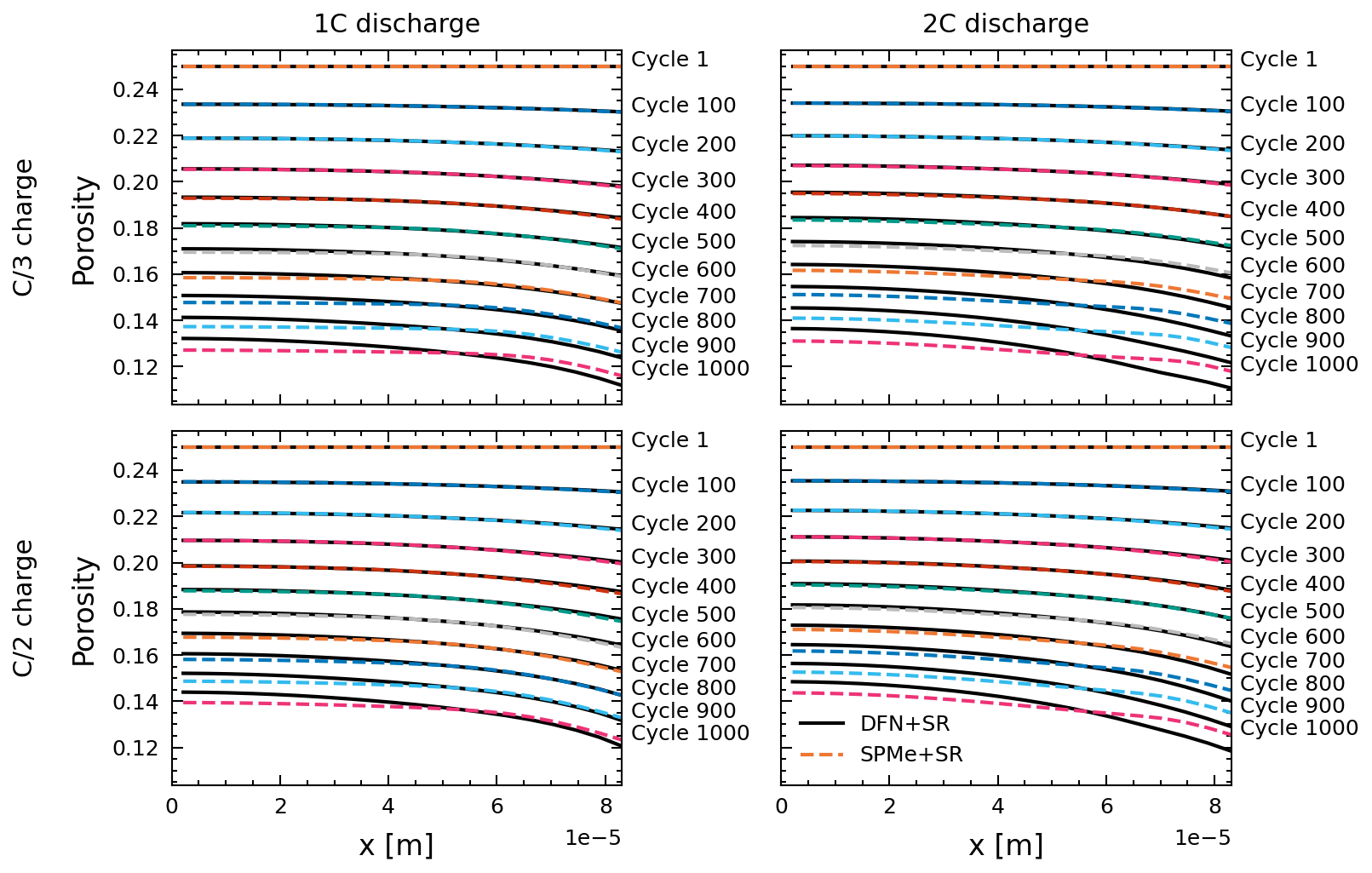}
    \caption{Compare porosity for the model with SEI growth and porosity change.}
\end{figure}

\begin{figure}[!htb]
    \centering
    \includegraphics[scale=1]{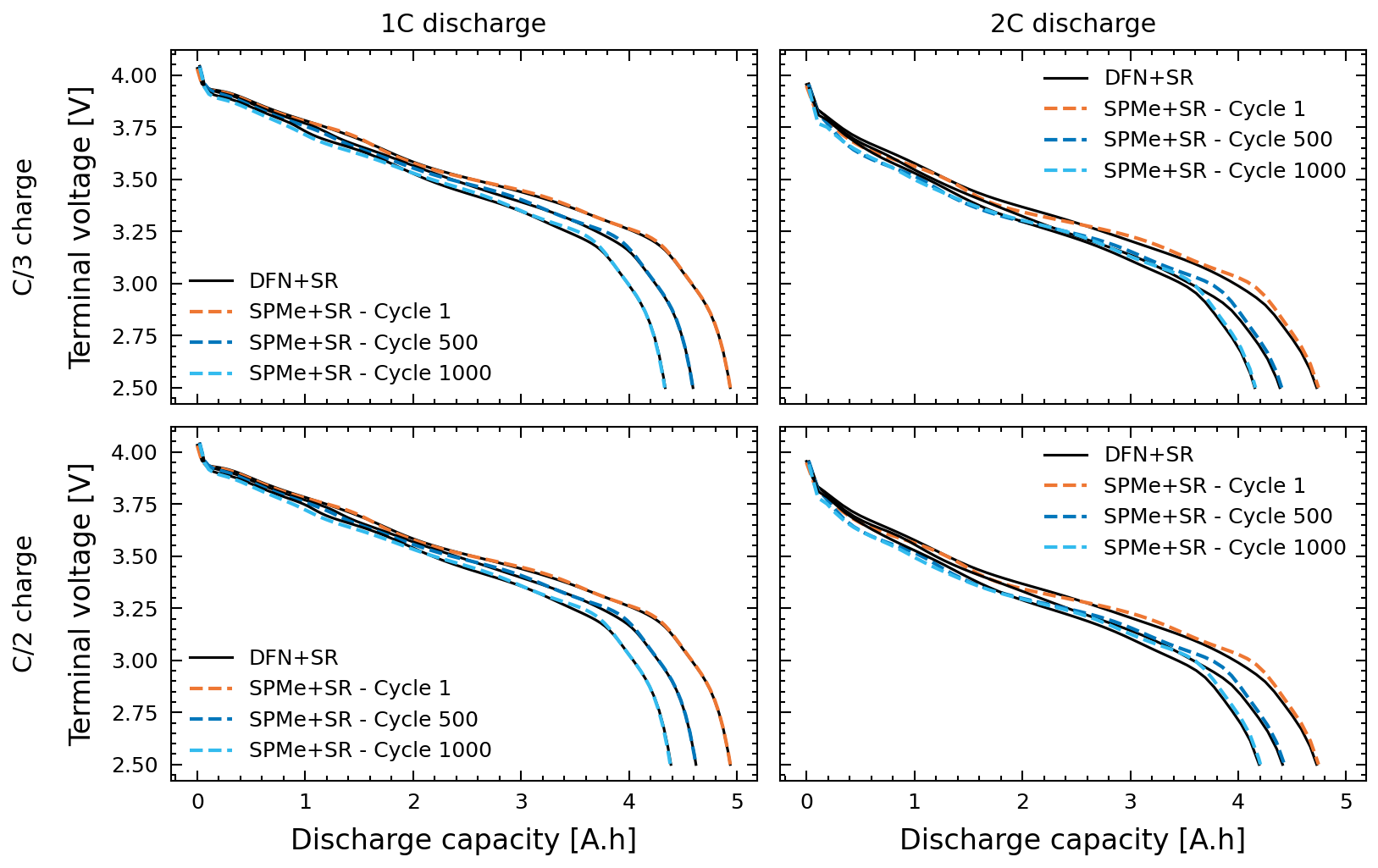}
    \caption{Compare voltage for the model with SEI growth and porosity change.}
\end{figure}



\FloatBarrier

\subsection{Lithium plating}
Next we show the results for lithium plating only.

\begin{figure}[htp]
    \centering
    \includegraphics[scale=1]{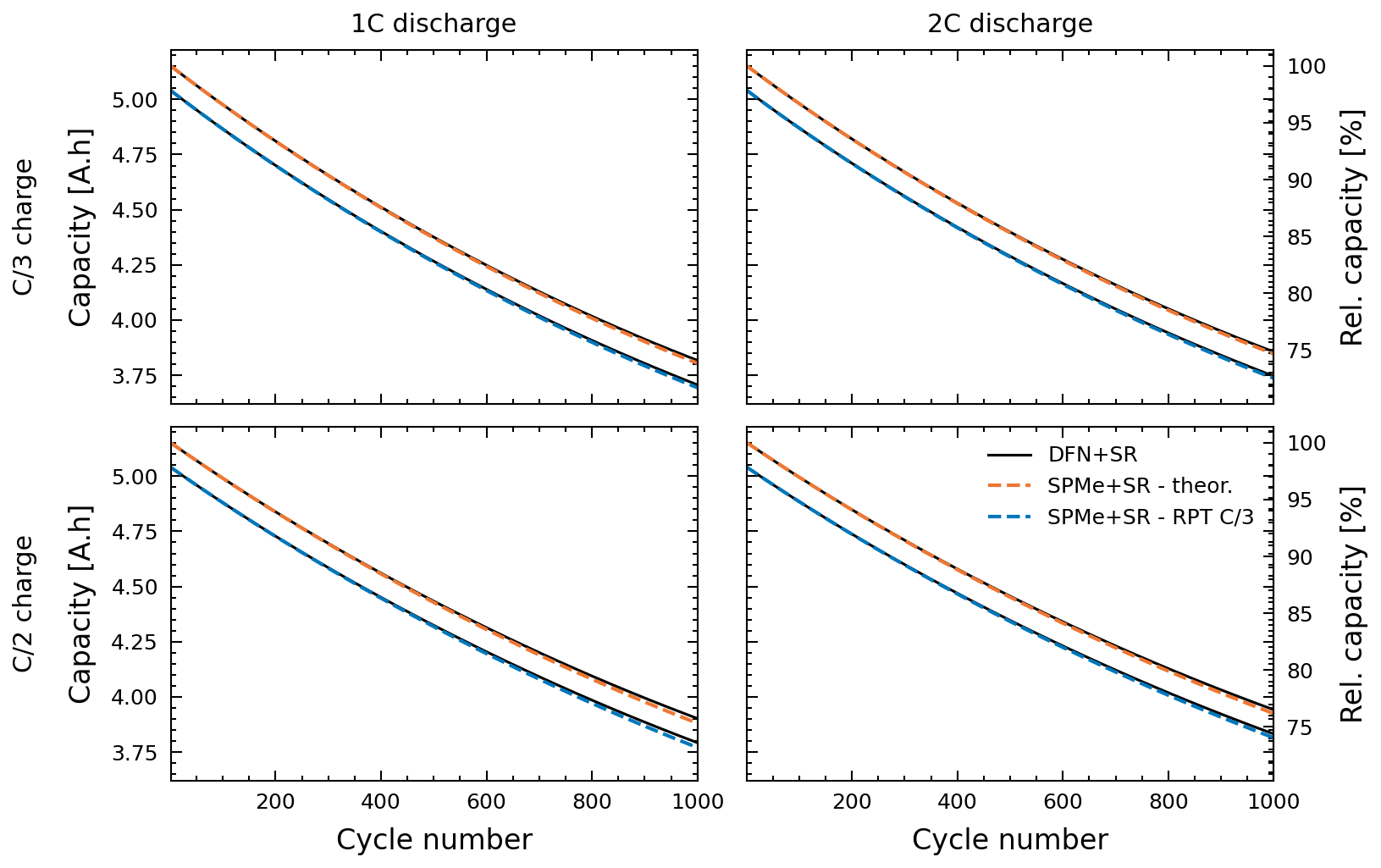}
    \caption{Compare capacity for the model with lithium plating and porosity change.}
\end{figure}

\begin{figure}[htp]
    \centering
    \includegraphics[scale=1]{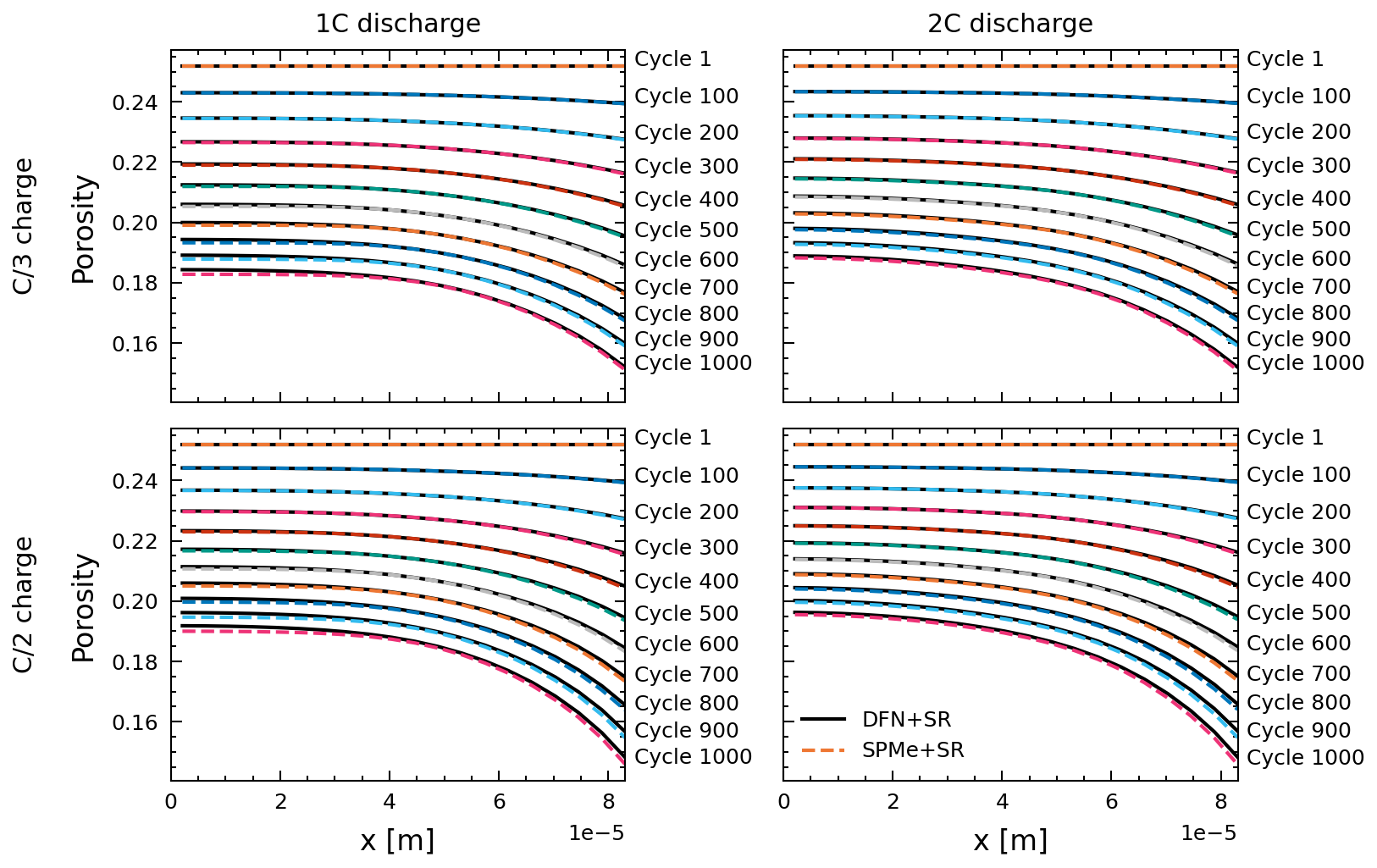}
    \caption{Compare porosity for the model with lithium plating and porosity change.}
\end{figure}

\begin{figure}[htp]
    \centering
    \includegraphics[scale=1]{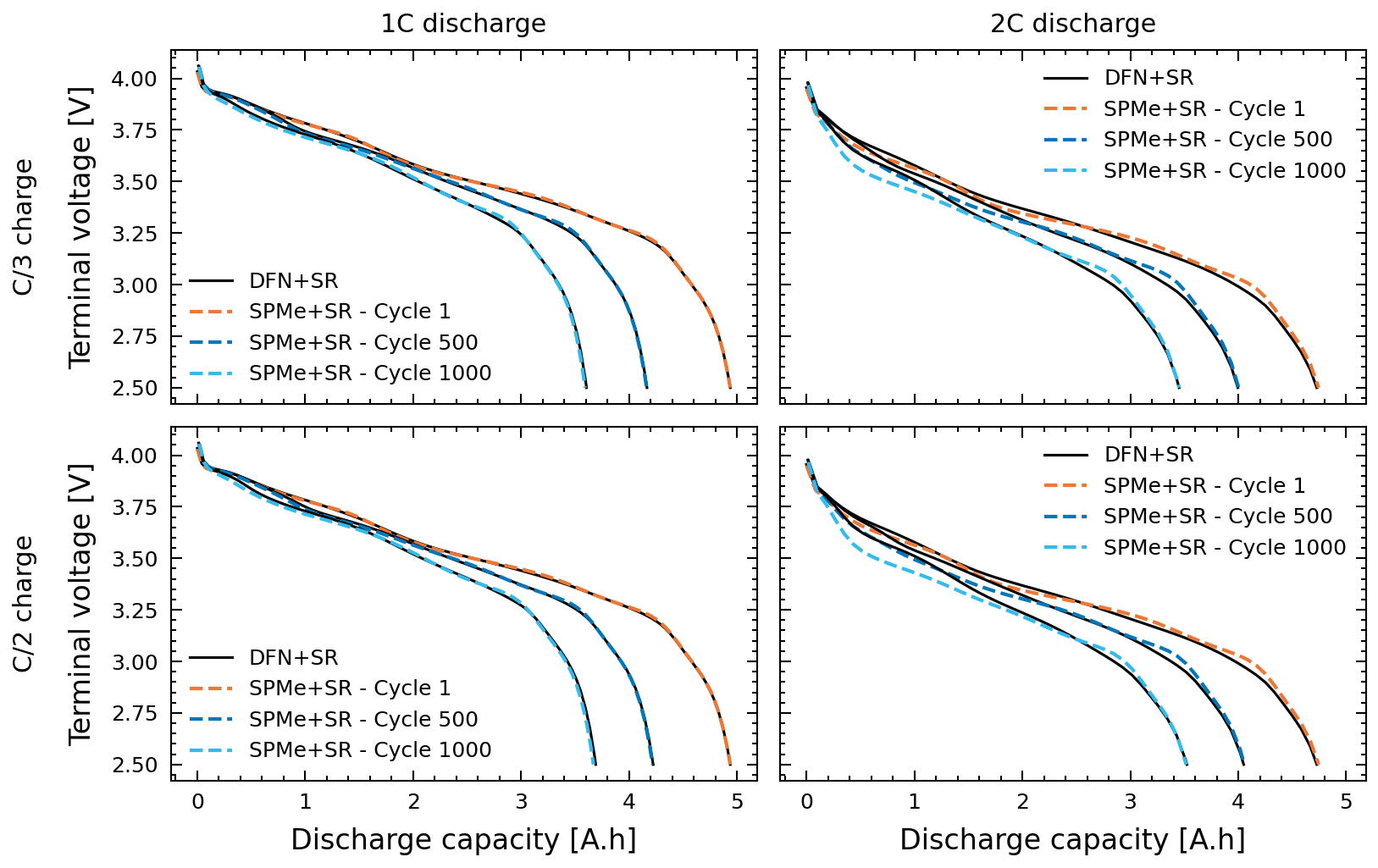}
    \caption{Compare voltage for the model with lithium plating and porosity change.}
\end{figure}

\FloatBarrier

\subsection{SEI growth and lithium plating}
Finally, we show the results for SEI growth and lithium plating combined.

\begin{figure}[htp]
    \centering
    \includegraphics[scale=1]{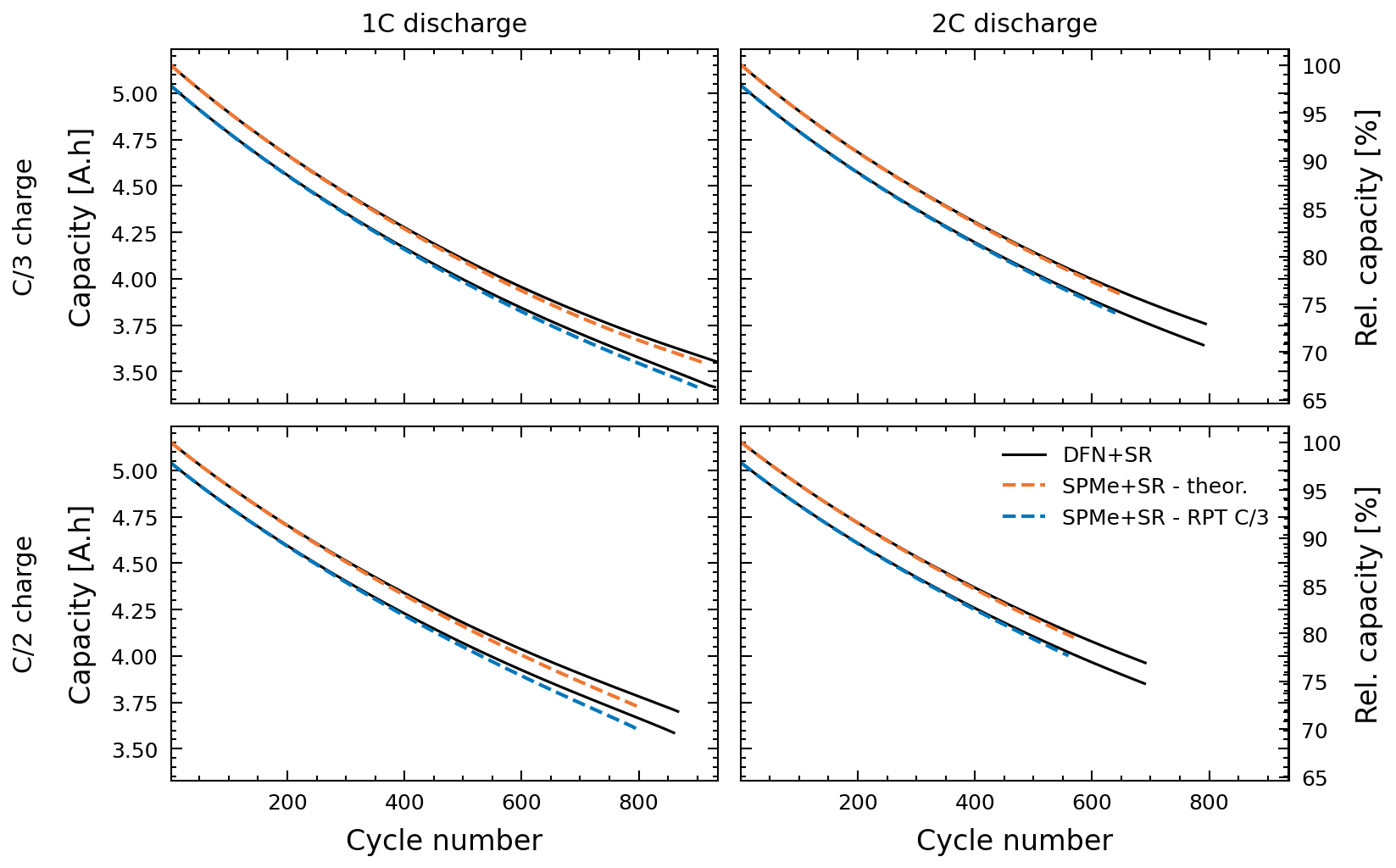}
    \caption{Compare capacity for the model with SEI growth, lithium plating and porosity change.}
\end{figure}

\begin{figure}[htp]
    \centering
    \includegraphics[scale=1]{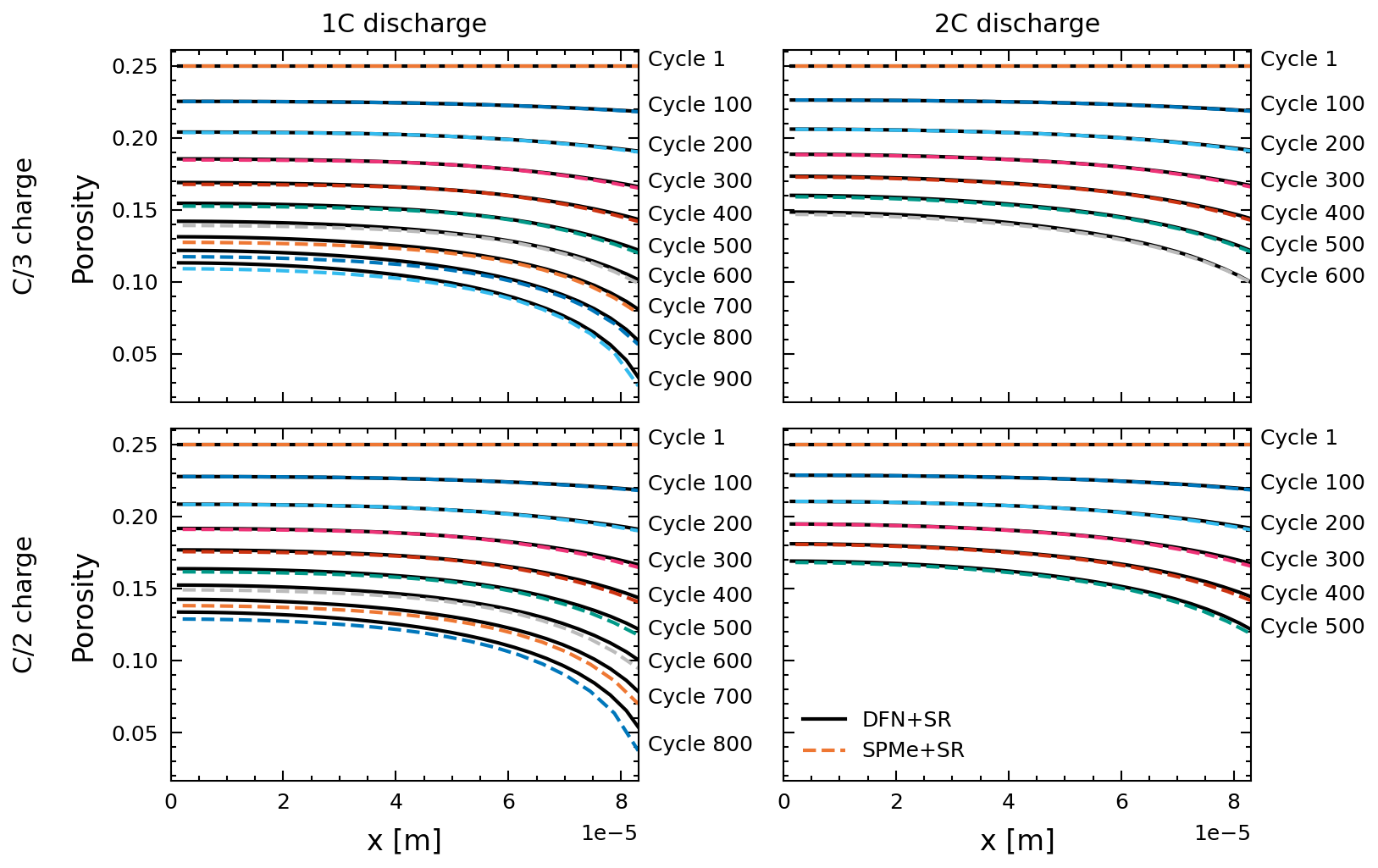}
    \caption{Compare porosity for the model with SEI growth, lithium plating and porosity change.}
\end{figure}

\begin{figure}[htp]
    \centering
    \includegraphics[scale=1]{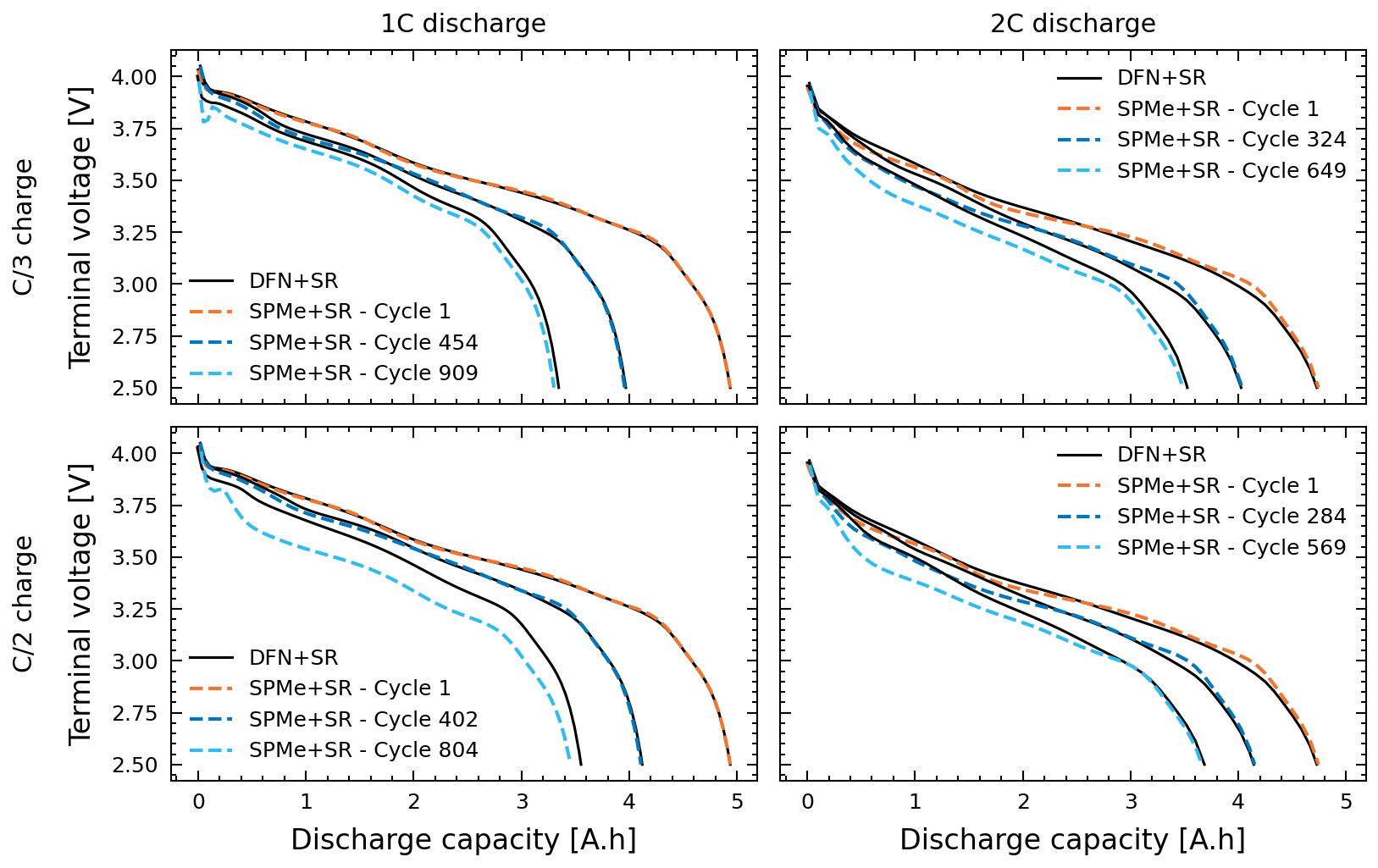}
    \caption{Compare voltage for the model with SEI growth, lithium plating and porosity change.}
\end{figure}

\FloatBarrier

\section{Model checklist}

\begin{table}[!h]
    \begin{tabular}{@{}p{2in}p{4.5in}@{}}
        \toprule
        Manuscript Title: & 
        Systematic derivation of a Single Particle Model with Electrolyte and Side Reactions (SPMe+SR) for degradation of lithium-ion batteries 
        \\ \midrule
        Submitting Author${}^{\ast}$: & 
        Ferran Brosa Planella
        \\ 
    \end{tabular}
    
    \begin{tabular}{@{}lp{5.5in}c@{}}
        \toprule
        \# & Question & Y/N/NA${}^{\dagger}$ \\ \midrule
        1 & Have you provided all assumptions, theory, governing equations, initial and boundary conditions, material properties (e.g., open circuit potential) with appropriate precision and literature sources, constant states (e.g., temperature), etc.? & 
        Y
        \\
        & \multicolumn{2}{p{6.3in}}{\textbf{Remarks:}
        The code is publicly available.
        } \\ 
        
        \midrule
        2 & If the calculations have a probabilistic component (e.g. Monte Carlo, initial configuration in Molecular Dynamics, etc.), did you provide statistics (mean, standard deviation, confidence interval, etc.) from multiple $\left( \geq 3 \right)$ runs of a representative case? & 
        NA
        \\
        & \multicolumn{2}{p{6.3in}}{\textbf{Remarks:}
        The article does not include any probabilistic calculations.
        } \\ 
        
        \midrule
        3 & If data-driven calculations are performed (e.g. Machine Learning), did you specify dataset origin, the rationale behind choosing it, what all information does it contain and the specific portion of it being utilized? Have you described the thought process for choosing a specific modeling paradigm?  & 
        NA
        \\
        & \multicolumn{2}{p{6.3in}}{\textbf{Remarks:}
        The article does not include any data-driven calculations.
        } \\ 
        
        \midrule
        4 & Have you discussed all sources of potential uncertainty, variability, and errors in the modeling results and their impact on quantitative results and qualitative trends? Have you discussed the sensitivity of modeling (and numerical) inputs such as material properties, time step, domain size, neural network architecture, etc. where they are variable or uncertain? & 
        Y
        \\
        & \multicolumn{2}{p{6.3in}}{\textbf{Remarks:}
        The details of the numerical methods used have been provided.
        } \\ 
        
        \midrule
        5 & Have you sufficiently discussed new or not widely familiar terminology and descriptors for clarity? Did you use these terms in their appropriate context to avoid misinterpretation? Enumerate these terms in the `Remarks'. & 
        Y
        \\
        & \multicolumn{2}{p{6.3in}}{\textbf{Remarks:}
        The acronyms appearing in the article have been described the first time they appear. These include: Single Particle Model with electrolyte (SPMe), Doyle-Fuller-Newman model (DFN), solid-electrolyte interphase (SEI), ordinary differential equations (ODEs), differential-algebraic equations (DAEs).
        } \\ \bottomrule
    \end{tabular}
    
    {
    \footnotesize
    ${}^{\ast}$ I verify that this form is completed accurately in agreement with all co-authors, to the best of my knowledge.
    }
    
    {
    \footnotesize
    ${}^{\dagger}$ Y $\equiv$ the question is answered completely. Discuss any N or NA response in `Remarks'.
    }
\end{table}